\documentclass[a4paper,onecolumn,11pt]{article}
\pdfoutput=1
\usepackage[utf8]{inputenc}
\usepackage[english]{babel}
\usepackage[T1]{fontenc}
\usepackage{amsmath}
\usepackage{hyperref}

\usepackage{tikz}  
 
\usepackage{fullpage}
\usepackage{amsfonts,amssymb}
\usepackage{amsthm}
\usepackage{tikz}
\usetikzlibrary{calc}
\usetikzlibrary{arrows}
\usepackage{tikz-3dplot}
\usepackage{color}  
\usepackage{hyperref}
\usepackage{bm}
\usepackage{float} 
\usepackage{appendix} 
\usepackage{lscape}
\usepackage{cite}
\usepackage{mathrsfs}
\usepackage{appendix}
\usepackage{setspace}
\usepackage{mathtools}
\usepackage{authblk} 

\theoremstyle{definition}

\begin{document}
	
	\title{Watching the Clocks: Interpreting  the Page-Wootters Formalism and the Internal Quantum Reference Frame Programme}
	\date{\today}

 \author[1]{Emily Adlam} 
	\date{\today} 
	\affil[1]{University of Western Ontario}

	\maketitle
	
	\tableofcontents
	
	\newpage
	
	\section{Introduction}
	
Recently the field of quantum foundations  has seen a surge of interest in the Page-Wootters (PW) formalism, as well as the development of a new research programme based on the related notion of an internal quantum reference frame (IQRF). A number of   exciting results have come out of this research, giving new insight into deep questions around issues like the nature of time, reference frames, and the equivalence principle. These concerns put the PW and IQRF research programmes right at the interface of the foundations of quantum mechanics and the ongoing search for a theory of quantum gravity, so it is of great interest to understand what exactly   the results of these  programmes mean for our understanding of these fields. In this article we aim to clarify the foundational consequences of some of the main themes of the PW and IQRF programmes, although of course we will not be able to cover everything that researchers in these fields have have achieved.

One feature of these research programmes which raises a number of questions is the fact  that the PW formalism, and the IQRF research programme more generally, don't  obviously seem to make room for a mechanism of wavefunction collapse or any other such prescription which would ensure that measurements have unique outcomes. So one might think that in order to take these research programmes seriously we will have to adopt the Everett interpretation, a neo-Copenhagen interpretation, or some other such interpretation which does not insist on unique measurement outcomes.  Thus in order to understand  the foundational significance of this research, it is important to establish whether or not the PW and IQRF formalisms implicitly depend on a certain interpretation of quantum mechanics, and whether there is any possibility of making sense of their results in the context of a single-world realist interpretation. 
 
There are also important questions surrounding the operational credentials of the PW and IQRF formalisms. It is common for proponents of both approaches to motivate  their research in operational terms - for example, ref \cite{2020cqtd} argues that `\emph{Extending the operational view to quantum theory, one is led to define time through measurements of quantum systems serving as clocks.'} This operational approach to time sounds very reasonable (and indeed it is inherited from Einstein's approach to time in special and general relativity), but it's important to keep in mind that these frameworks do not often explicitly model observers and  therefore there remains a gap to be bridged between the formal results and the operations performed by actual observers. Closing that gap will likely require us to adopt some standpoint on the role of observers and the nature of the probabilities that the PW and IQRF formalisms ascribe to measurement outcomes. 
 
Thus after giving a brief introduction to the PW and IQRF research programmes, we will divide our enquiry up into four   questions, which in fact will all turn out to be interrelated: 

\begin{enumerate}

\item What account does the PW formalism give of the nature of time? 

\item What do the probabilities predicted by the PW formalism refer to?
  
\item Is a `quantum reference frame' really a reference frame in an operationally meaningful sense?  
 
\item What does it mean to say that superposition is relative to a reference frame, and what consequences does that have for the interpretation of quantum mechanics? 
 
\end{enumerate} 

We will discuss some specific difficulties that arise in attempting to interpret the PW formalism in particular, and then suggest one possible interpretational approach based on an approach  advocated in other contexts in refs \cite{2007mep,Kent,2015KentL}. This `final-measurement' interpretation demonstrates that it may be possible to make sense of the PW formalism within a single-world realist context; it preserves all of the mathematical results of the PW and IQRF research programmes, but does alter the conclusions we can draw from some of these results. It is not our intention to claim that the final-measurement interpretation  is the only possible interpretation of the PW and IQRF formalisms, but the broader point we wish to make is that \emph{some} interpretation of the framework is necessary if we are to use it to make foundational claims, and moreover the choice of interpretation will inevitably  change our understanding of some of the mathematical results. 

We demonstrate this by arguing  that if reference frames are required to have operational significance such that they can be associated with measurement procedures, different interpretations of quantum mechanics will lead to different conclusions about what constitutes a quantum reference frame. Similarly, we argue that the notion of superposition should take into account the way in which an instantaneous state is embedded in ongoing dynamical evolution, which leads to a more nuanced way of thinking about the relativity of superposition.  Thus this paper serves to emphasize the fact that  conclusions   drawn from mathematical frameworks like the PW and IQRF formalisms will not typically be interpretation-neutral, so it is important for research programmes addressing foundational issues to explicitly address questions about the interpretation of quantum mechanics.

\section{Background}

\subsection{Internal Quantum Reference Frames \label{intro} } 
 
Typically in physics we describe the behaviour of systems relative to some fixed reference frame which is understood as being external to the systems we aim to describe. However, there are presumably no reference frames external to the whole universe, so ultimately we must always define the behaviour of subsystems of the universe relative to other subsystems of the universe. Moreover, reference frames are often treated as abstract mathematical constructions in physics, but taking this kind of internal, relative view requires us to decide what the physical correlate of a reference frame is.  The  Internal Quantum Reference Frames (IQRF) programme\cite{2019cov,2020qctloe,2020acop,2020htsbrqc,giacomini2021quantum} offers one possible answer to this question: the physical correlate of a reference frame is simply an individual physical system. Thus to arrive at a reference frame, we fix a reference  system and  rewrite our models such that \emph{`the position and momentum of the reference frame are not dynamical variables when considered from the reference frame itself  ...  the reference frame is not a degree of freedom in its own description, but external systems to it are.'}\cite{2019cov}  The IQRF programme  studies the problem of establishing and switching between such internal reference frames. A variety of techniques have been applied to this problem, but the field seems to have largely converged on using the framework of  constraint quantisation first formulated by Dirac\cite{dirac2019principles}. 
 
In constraint quantisation, we start with a `kinematical Hilbert space' that defines all the possible quantum degrees of freedom, and then we impose constraints encoding the idea that real physical states should be  invariant under certain transformations deemed to be physically meaningless - for example, diffeomorphisms and global time translations of the whole universe\cite{Kucha1993CanonicalQG}. Mathematically this is done by requiring that a certain set of constraints should vanish on allowed states: for example, since we require that physical states should be invariant under translations of the whole universe, the energy of the whole universe obeys a superselection rule expressed by the constraint $H  || \Psi \rangle = 0$, where $H$ is the Hamiltonian of the whole universe and  $|| \Psi \rangle$ is the universal quantum state. This constraint is known as the Hamiltonian constraint, and its constraint equation is known as the Wheeler DeWitt equation\cite{2015seqg}. The solutions to this equation  may not be normalizable with respect to the inner product defined on the kinematical Hilbert space, so we introduce a new inner product which is used to complete the solution space of the constraint equation, giving us what is known as the `physical Hilbert space,' to reinforce the fact that only states satisfying all the constraints are considered physically possible. Similarly, only operators which commute with the constraints on the physical state space are considered to be genuinely observable, and  such quantities are known as Dirac observables\cite{hoehn2022quantum}.

In this framework, the fact that the universe is constrained to lie in a subspace of the full kinematic Hilbert space entails that there is redundancy associated with standard descriptions of physical systems relative to a fixed background coordinate system. The IQRF research programme seeks to  remove this redundancy  (i.e. to perform gauge-fixing) by relativizing such descriptions  to some chosen physical system. To see how this works, consider a particle $P$ and system $S$ which are in some joint state $\psi$, and suppose that the system as a whole is invariant under symmetry transformations belonging to some group $G$ (e.g. diffeomorphisms and/or global time translations), so any member of the equivalence class of states related by transformations in $G$ is a legitimate description of the composite system. We can remove the redundancy  by choosing to describe the system relative to the fixed reference frame of $P$: we simply look  for a state $\psi'$ which is related to $\psi$ by a symmetry transformation $G$ such that $\psi'$ can be written in the product state $|e\rangle_P \otimes \phi_S$, where $|e\rangle_P$ is the unit element of the symmetry group and $\phi_S$ is any state of $S$: then we say that $\phi_S$ is the state of $S$ relative to $P$, or in the reference frame of $P$\cite{2021mmetc}.  States $\psi$ for which this transformation is possible are sometimes known as `alignable,' and in the case where the transformation is possible it is unique up to a global phase.  Transformations between such reference frames are elements of $G_{sym}$, which includes the transformations in $G$ and also the additional set of transformations obtained by selecting a transformation in $G$ in a way that depends on gauge-independent relative facts about the configuration of the system\cite{2021mmetc}. That is to say, transformations between reference frames are symmetry transformations.

We note that there are two different ways of interpreting this procedure. The kinematical Hilbert space is the set of states distinguishable relative to an external reference frame, while the physical Hilbert space is the set of states distinguishable from the point of view of an observer inside the universe. So one option is to take a realist approach to the whole kinematical Hilbert space and thus suppose that transformations like global time translations of the whole universe are in principle possible, although they would not have any effects that would be observable from inside the universe. For example, this would be the appropriate approach to adopt if we were dealing with a constrained system constructed in a lab, in which case the laboratory provides the external reference frame and thus translations of the whole system are physically meaningful.  The alternative is to take a deflationary approach to the kinematical Hilbert space where we hold that there is no reference frame outside of the universe as a whole and thus transformations like  global time translations of the whole universe are physically meaningless (assuming  that the universe has no boundaries). From this point of view, the kinematical space is not literally a set of possible states for the universe; it should simply be regarded as a mathematical abstraction which helps us construct the true set of possible states for the universe, i.e. the physical Hilbert space.

Recent work\cite{delahamette2021perspectiveneutral,2019switch} has shown that the set of all internal quantum reference frames can be unified by a `perspective-neutral' universal quantum state defined on the physical Hilbert space of gauge-invariant states. We can obtain `quantum coordinate maps' which take us from the universal quantum state into the perspective of an individual quantum reference system, and then  in order to transform between reference frames we can always transform  to the perspective-neutral universal state using the inverse of the quantum coordinate map, and then apply another quantum coordinate map to arrive at a different reference frame. Gauge-invariant relational observables can be constructed in this framework to encode physical properties relative to an individual reference frame, so the perspective-neutral framework still upholds the basic idea that ultimately all physical quantities in our universe are relational.  

In appendix \ref{others} we explain the relation between the IQRF formalism and several other relevant research programmes dealing with quantum reference frames. 

\subsection{The Page-Wootters Formalism}

The internal reference frame approach is closely related to the Page-Wootters (PW) formalism\cite{2021gpr}, which was originally proposed by Page and Wootters\cite{PhysRevD.27.2885} as a response to the so-called `problem of time' that follows from the constraint quantisation procedure\cite{articleIsh}: as a result of the constraint equation  $H || \psi \rangle = 0$,  all Dirac observables must commute  with $H$ and therefore no observable quantities in the theory can have any time-dependence. That is to say, in this picture time itself is treated as a form of gauge.   This may sound troubling, and has been questioned by several authors on the grounds that states of the world at different times are obviously physically distinct\cite{doi:10.1142/S0218271811019347}. However, this feature makes sense if one regards the state  $ || \psi \rangle $ as a  global description which represents the whole history of the universe, as naturally a representation of the whole of history cannot undergo any change. Moreover, because  we are dealing with a theory which is deterministic in both directions of time, the  state at any one time determines the whole course of history, so there is a sense in which states at different times are  genuinely  physically equivalent in the sense that they can be regarded as different descriptions of one and the same global configuration $ || \psi \rangle$.  The `problem of time' then reduces to the problem of explaining how it is that observers within this block universe have temporal experiences.

%'The theory provides us with only a timeless wave function on the configuration space of 3-metrics, i.e. on the possible configurations of space, not of spacetime. But how can a theory that provides us (at best) with a single fixed probability distribution for configurations of space ever be able to describe the always changing world in which we live? This, in a nutshell, is the problem of time in canonical quantum gravity.'
%Goldstein and Teufel
%So not sure the above is quite right - because we are looking at a probability distribution over 3-metrics, so what does it even mean?

The PW formalism aims to answer this question by employing   an internal reference frame, as in the IQRF programme; specifically,  we choose a reference frame that can be regarded as a clock, such that we can understand time in terms of states of the universe relative to different clock readings. To do this, we choose a subsystem $C$ of the universe to act as a clock and then define a `time observable' $E_C(t) = |t \rangle \langle t |$ which acts on the clock system $C$ and which transforms covariantly with respect to the group generated by the Hamiltonian $H_C$ of that subsystem, i.e. $E(t + t') = e^{-iH_C t'} E(t) (e^{-iH_C t'})^{\dagger}$. For simplicity it is common to work largely with `ideal clocks'  in which the time observable is sharp, i.e. the clock states are orthogonal, though it should be recognised that most real clocks are unlikely to have this feature; non-ideal clocks can also be used in the PW framework but in that case we must take a little more care in defining the covariant time observable.   Using the covariant time observable and the universal quantum state $|| \Psi \rangle$ we can define a state for the rest of the world,  $W$, conditional on $C$ reading time $t$: $| \psi_W(t) \rangle = \langle t | \otimes I_W || \Psi \rangle$.  We define an inner product by requiring that this relative state always has norm one. Finally, by differentiating with respect to time and invoking the Hamiltonian constraint, we infer that $i \frac{d  }{dt} | \psi_W(t) \rangle = H_S | \psi_W(t) \rangle$, i.e. relative to $C$ the system $W$ undergoes the standard Schrodinger evolution\cite{2020cqtd}. In essence, this works because the constraint $H || \Psi \rangle$ entails that the total energy is conserved and therefore since the time variables are defined in terms of the Hamiltonians encoding the energy, it follows that if the time variable for one system changes by $+t$ then the other variable changes by $- t$, so the clock and the system are necessarily correlated. According to the PW approach, these sorts of relational facts are supposed to be \emph{constitutive} of time - that is, absolute time has been eliminated in favour of a relational picture.

\section{The Page-Wootters account of time}

The principal foundational claim of proponents of the PW formalism is that it  provides a novel account of time suitable for the prima facie `timeless' setting of canonical quantum gravity, so an obvious starting point for our enquiry is to ask what precisely is the content of that account. To make things more concrete, consider the following  question:  given a state of the world like $  | t \rangle_C \otimes | \psi \rangle_S$, what is it that makes it the case that this is the correct state, rather than some other state  $| t \rangle_C \otimes | \psi' \rangle_S$? This question might seem trivial, but keep in mind that if we were working in the context of standard quantum mechanics,  the link between $C$ and $S$ which determines that the correct state is $  | t \rangle_C \otimes | \psi \rangle_S$ rather than $  | t \rangle_C \otimes | \psi' \rangle_S$ is of course the fixed background time metric:  system $S$ is in the state $| \psi \rangle$ and not $| \psi' \rangle$ \emph{at the same time} as system $C$ is in the state $ | t \rangle_S$. But   in the PW formalism we are not allowed to appeal to a fixed background time metric, so we need some alternative way of justifying the claim that the correct state is $  | t \rangle_C \otimes | \psi \rangle_S$ and not  $  | t \rangle_C \otimes | \psi' \rangle_S$ -  that is to say, we need an alternative account of the physical meaning of the tensor product which doesn't require a fixed background time metric. If such an explanation can't be given,  then we may have to conclude that the account of time offered by the PW formalism in fact derives its plausibility from  intuitions developed in the context of standard quantum mechanics which are no longer legitimate in the absence of a fixed background metric, and thus this account may not offer a satisfactory solution to the problem of time in the absence of that background temporal metric. 

One common way of answering this question is to suggest that the PW approach shows us  that the   temporal relations between clock and system are based on  \emph{entanglement}\cite{2014fmqe, PhysRevD.92.045033,PhysRevD.95.043510}.  Prima facie this claim seems surprising, because macroscopic objects do not typically exhibit entanglement, so it would seem that entanglement  should be irrelevant to the problem of making sense of our macroscopic temporal experiences.  But in fact, as explained in ref \cite{2021trinity} the relevant entanglement between clocks and systems is not the usual sort of entanglement   which we can detect by measurement. The entanglement in question is defined  on the  kinematical Hilbert space, not the physical Hilbert space - and kinematical entanglement can't be probed using gauge-invariant Dirac observables, which means that it can't be detected from within the universe. Therefore it could be the case that clocks and systems are generically  entangled in the kinematical Hilbert space, even though we can't observe that entanglement by means of any measurements. So one might try to argue that despite being unobservable,  kinematical entanglement plays some kind of coordinating role and thus is ultimately responsible for the relationships between clocks and systems expressed by states like $  | t \rangle_C \otimes | \psi \rangle_S$.  

However, recall that according to the second interpretation of the constraint framework discussed above, the kinematical Hilbert space is considered to be not only unobservable but \emph{unphysical} - it does not represent a real state space, just a mathematical abstraction. This interpretation of the constraint framework is quite common in the literature and evidently adherents of it have good reason to question the idea that entanglement in the kinematical Hilbert space is a crucial feature of the relationship between clocks and systems, since entanglement in a wholly abstract space presumably cannot lead to substantive consequences like our experiences of time. It should also be noted that the physical Hilbert space does not in general inherit the  tensor product structure of the kinematical Hilbert space\cite{hoehn2022quantum}, so the decomposition into `clock' and `system' may not even exist in the kinematical Hilbert space. In general, if the kinematical Hilbert space has two subsystems $A, B$, when we transform to the physical Hilbert space it is possible that we will end up with a `clock' made up of some of the degrees of freedom of $A$ and some of $B$, and likewise a `system' also made up of some of the degrees of freedom of $A$ and some of $B$. Moreover we also generally find that tensor product decompositions change when transforming between different internal references frames, so it may also  be the case that the tensor product structure of the kinematical Hilbert space is preserved in some possible internal references frames but is not preserved in others. So generically it will not be correct to say that a  `clock' and  `system' in the physical Hilbert space are entangled in the kinematical Hilbert space, because they may not even exist as distinct subsystems in the kinematical Hilbert space.

Moreover,  there are a number of formalisms which achieve a relational description very similar to the PW approach without employing entanglement at all. For example, ref \cite{2021trinity}  has shown that  one can also obtain the same relational dynamics without even postulating any \emph{kinematical} entanglement between clock and system degrees of freedom: first we employ the Heisenberg trivialisation which transforms the clock and system state into a product state (see appendix \ref{others}) and only after that do we employ the PW reduction scheme. So it certainly can't be kinematical entanglement which plays the role of linking clock and system within this particular version of the PW approach. Likewise, Poulin\cite{2006poulin} has shown that if we employ statistical group averaging (rather than the coherent group averaging that features in the usual PW approach) we arrive at a description very similar to the PW account except that  the universal state is mixed and thus the correlations between  clocks and systems are entirely classical. And Vedral\cite{vedralcewe} has recently set out a classical version of the Page-Wootters construction and shown that `\emph{the time variables of the two subsystems are correlated, not because the subsystems are coupled, but because the overall energy is conserved.}' Vedral's account requires us to suppose that the classical state of the whole universe is mixed, and in a classical setting a mixed state must be understood either as a characterisation of our  lack of knowledge or as a description of an ensemble of universes, so it's unclear that this picture is compatible with the standard classical picture of a single universe with a well-defined state, but  nonetheless it does demonstrate that the conservation of energy is already enough to derive relationships between time variables without any need to appeal to entanglement. These frameworks show  that we can get  relational dynamics very similar to the PW picture   without any   entanglement at all, kinematical or otherwise, and thus they seem to undermine the argument that  kinematical entanglement is playing some crucial role in the account of time offered by the PW formalism.

So if it's not entanglement which  links the state of the clock to the state of the system in the PW formalism,  what \emph{is} the nature of this link? Another popular idea is that the PW formalism has reduced time down to \emph{correlations} between the states of clocks and the states of systems: for example, ref \cite{2020qctloe} presents `\emph{a “timeless” approach according to which time emerges though correlations between “what a clock shows” and the state of the system.}'  But there are several obstacles to making physical sense of these sorts of correlations in the PW framework. First,  in order to observe correlations between events we must already be provided with a way of putting those entities into pairs - for example, we can't observe Bell correlations if we are simply given a set of 1000 distinct measurement settings and the corresponding outcomes, we must also be told which pairs of measurements belong together, i.e. are associated with the same Bell pair. So in the PW case, how are we supposed to identify pairings of a `state of a clock' and a `state of the rest of the universe,' in order to observe these supposed correlations? The literature on the topic frequently uses locutions like  `\emph{given that clock A shows a time $t$}'\cite{2020qctloe} or `\emph{$W$ measures outcome w at clock time $t_2$}'\cite{2021gpr} or `\emph{the probability that one clock reads a given proper time conditioned on another clock reading a different proper time},'\cite{2020cqtd} but all of these phrases seem a little suspect, because in ordinary language they   have a temporal connotation:  `the state of the system given that the clock reads $t$' usually means something like `the state of the system \emph{at the same time as} the clock reads $t$.' But if we are hoping to \emph{reduce} time to correlations then of course we can't understand phrases like `given that,' `at,' `conditional on' and so on in this temporal sense, and yet the literature does not  obviously seem to offer any alternative way of understanding them. So in order to interpret the account of time offered by the Page-Wootters formalism in terms of correlations we will need to come up with some more concrete way of saying what it means to condition the outcome of one measurement on another in a context where we are not able to appeal to a temporal background metric. 

Second, we typically think of correlations as being exhibited between \emph{events} such as measurement outcomes and clock readings, and it's a little unclear what the `events' are in the PW picture. Ref \cite{2020qctloe}  offers `\emph{a local, operational definition of an “event”, in which an operation is applied to a quantum system conditioned on a clock reading a specific “time”.}  However, events of this kind are not in and of themselves adequate to play a role in an account of our temporal experiences, because the IQRF formalism typically avoids explicitly modelling observers and hence the formalism does not describe events of this kind as actually being observed. Nor is it straightforward to say what it would mean for these events to be observed, because the relative state $| \psi_W(t) \rangle = \langle t | \otimes I_W || \Psi \rangle$  is a state of the whole world minus a single clock, so real observers will always be subsystems of the `rest of the world,' which means we can't model their observations using the usual quantum  measurement formalism which presumes external observers. So in order to translate the PW approach into statements about the temporal experiences of actual observers, we will need some prescription about how the state $| \psi_W(t) \rangle$ relates to observations made by observers who are subsystems of $W$. Moreover, even if we can find a way to identify a representation of an observer having some experiences within the state $| \psi_W(t) \rangle$, it doesn't seem obvious that we can conclude that the observer  is having that experience \emph{at time $t$}, because the state $|\psi_W(t) \rangle$  doesn't include a representation of  the observer, or indeed anyone else, measuring the clock and getting the result $t$. Indeed  it is common for applications of the PW formalism to represent the clock as non-interacting, and when interactions are considered they are usually gravitational interactions\cite{2019quantizinginterac}, so in most of this research there is no provision made for observers to actually obtain time readings from the clock. Some such provision must be made if we are to understand how the PW account of time in terms of correlations can explain our temporal experiences. 

Third, in order to define correlations between clocks and systems we obviously have to calculate some  probabilities (or at least relative frequencies). Usually the probabilities prescribed by the IQRF formalism are for measurement outcomes conditioned on various clock readings; but if these measurements are actually performed  and the standard formalism of wavefunction collapse is applied, the collapse mechanism would take the state outside the physical Hilbert space\cite{Kucha1993CanonicalQG}, so prima facie it seems like the IQRF formalism must forbid any such measurements to actually be performed. Page himself observed that the collapse of the wavefunction is only \emph{`part of one's computational process, not as a physical process, so it should not be surprising that the result is rather unphysical,'}\cite{PhysRevD.27.2885} but this response on its own doesn't seem entirely adequate - we may well accept that the collapse of the wavefunction is just a way of computing probabilities, but what exactly are these probabilities \emph{for} if not for unique measurement outcomes as defined by a wavefunction collapse? Without a mechanism for selecting a unique measurement outcome we seem obliged to conclude that \emph{all} of the outcomes of these potential measurements actually occur, which is problematic for the account of time in terms of correlations, as we can't define  any meaningful correlations between clock readings and outcomes if all of the outcomes always occur - the correlations will all be trivially equal to one. 

One way to address this issue would be to move to a version of the Everett interpretation\cite{Wallace}  and employ one of the strategies used by Everettians to make sense of probabilities in a context without unique measurement outcomes. However, the interpretation of probability is perhaps the most contentious feature of the Everett picture, and there remain serious doubts as to whether any of these strategies succeed\cite{AdlamEverett,Albert2010-ALBPIT,kent2009world}. We will not rehearse these debates here; suffice it to say that we are not convinced that the Everettian probability problem can be solved, and therefore we consider it important to establish whether the IQRF framework can be understood without appeal to the Everett interpretation  or similar. That is to say, we would like to know whether there is any way to extract a \emph{single-world realist} picture out of the IQRF framework - i.e. a picture in which measurements have unique outcomes and at least the macroscopic parts of reality exist in a single definite way. In such a picture, `correlations between clocks and systems' could be understood in a straightforward way by appeal to relative frequencies across the whole of history, and therefore we would have a clear physical interpretation for the account of time offered by the IQRF framework.\footnote{We also note that even if one is satisfied by the Everett account of probabilities, problems remain for an Everettian account of the PW formalism. It would still be necessary to add something to the PW picture representing some kind of interaction in which clocks are actually observed, and it would still be necessary to explain how to pair measurements on systems with measurements on clocks  in order to calculate correlations - the standard Everett interpretation assumes we have unitary evolution on a background spacetime, so in an Everettian version of the IQRF formalism we would have to take care to avoid illegitimately appealing to a background time metric which is no longer available. We will not discuss here how these problems might be solved in the Everett picture, but we would invite proponents of the Everett interpretation to consider the question.}

\section{The meaning of the PW probabilities}

Extracting a single-world realist picture out of the IQRF framework is not trivial, because if measurements have unique outcomes  then necessarily there exist some beables associated with those outcomes, and these beables will  single out some preferred `quantum reference frames' which represent the beables as being in sharp states in some physically relevant basis (e.g. in the literature, typically `beables' are expected to have sharp states in the position basis). Thus in order to obtain a single-world realist account must find some way of breaking the symmetry between quantum reference frames, which are all considered on an equal footing in the standard IQRF formalism. The symmetry breaking does not have to be based on any objective difference between IQRFs - it could simply be a random process akin to the collapse of the wavefunction. However it is not straightforward to employ a literal collapse of the wavefunction in the IQRF framework, because as noted, such a collapse would take the system outside of the physical Hilbert space. Moreover, since there is no universal background time in the IQRF framework, it would seem difficult to say anything sensible about the time at which the collapse of the wavefunction occurs if indeed it is supposed to apply across different reference frames - in particular, there will in general be no frame-independent way of determining the order in which measurements take place, and since changing the order of wavefunction collapses will typically change the predictions, it's hard to see how we could arrive at a frame-independent approach under these circumstances. 

 Ref   \cite{2019switch}  argues that the universal wavefunction in the perspective-neutral approach `\emph{encodes all the descriptions of the universe relative to all possible choices of reference system at once and constitutes the crucial link between these internal perspectives,}' so one might naturally hope to obtain a single-world realist description by collapsing the  universal wavefunction directly at the level of the physical Hilbert space in order to obtain a definite course of history. This collapse would not have to be thought of as literally occurring at any particular time; it would simply be understood as the process by which a course of history is selected and actualised all at once in an atemporal manner. However, there are several obstacles standing in the way of extracting concrete predictions out of this kind of collapse. The first obstacle is just the usual measurement problem in a new guise: we know that the collapse must resolve   macroscopic features of reality into a single definite history, but at the same time we know that it must leave some microscopic features of reality uncollapsed in order to allow characteristic quantum phenomena to occur for microscopic particles,   and quantum mechanics itself provides no precise prescription for where to put the line between the microscopic and the macroscopic other than the vague suggestion that it should be associated with measurements. So in order to make this work we would have to give some prescription in precise terms (i.e. not referring to vague, anthropocentric concepts like `measurement') to determine exactly which parts of the universal wavefunction are supposed to collapse in this atemporal collapse.

The second problem is that   the universal wavefunction is `\emph{without immediate physical interpretation}'  \cite{2019switch} - that is to say, because its primary role is to encode a large number of different perspectives,  it does not directly encode any definite histories which we could collapse to. If we tried to proceed by simply collapsing the wavefunction separately inside each perspective according to the probabilities given by the Born rule, then the outcomes of measurements in one perspective would not correspond in any systematic way to the outcomes of the same measurements viewed from other perspectives, and thus we would all be trapped inside our own perspective, unable to find out what other observers have witnessed. Not only would this lead towards solipsism, it would also prevent us from being able to obtain empirical confirmation for a theory like quantum mechanics which purports to describe not just a single perspective but reality as a whole, and therefore our  epistemic justication for believing quantum mechanics in the first place would be undermined. So it would seem that if the perspective-neutral collapse is to be made viable, some  additional structure will have to be added to the IQRF approach.

 We will return to the perspective-neutral approach shortly, but first let us   take a closer look at the way in which  probabilities are actually calculated in the Page-Wootters formulation.  This issue has given rise to a vigorous debate, as it was pointed out by Kuchar\cite{doi:10.1142/S0218271811019347} that since applying the effect operator for the measurement which is used to define the Page-Wootters probabilities throws the state out of the physical Hilbert space,   it appears that the formalism gives the wrong answer to two-time questions like `if I perform measurement $O_1$ at time $t_1$ and obtain outcome $X$, what is the probability that when I perform measurement $O_2$ at time $t_2$ I will obtain outcome $Y$?' \footnote{Of course, one possible way of responding to this problem is to move to a version of the framework which postulates neither probabilities nor relative frequencies - i.e. perhaps   regard the framework as simply a description of the  universal wavefunction, understood in something like an Everettian picture with no attempt to attach probabilities to branches. However, it would seem very hard to understand the physical meaning of the framework's predictions in the absence of any probabilities or relative frequencies, and without predictions which can be linked to observations it would seem impossible for us  to obtain evidence for or learn anything about the putative `universal wavefunction' in the first place, so this approach would lead to serious  epistemic problems similar to those which have been raised for more vanilla versions of the Everett interpretation\cite{AdlamEverett}.  Also, as noted above, we do need either probabilities or relative frequencies if we are to have correlations in the usual mathematical sense, so if the physical meaning of time in these frameworks is supposed to come down to correlations, then we don't have the option of simply refraining from ascribing probabilities.} There are  a few different ways of arriving at the correct two-time probabilities.   Dolby proposes  that we should apply the usual measurement operator, taking us out of the physical Hilbert space, and then immediately apply another operator which takes us back into the physical Hilbert space\cite{Dolby2004TheCP}.    Hoehn et al\cite{2021trinity} have shown that the correct two-time probabilities can be obtained as manifestly gauge-invariant relational Dirac observables within the constraint picture - this approach actually turns out to be formally equivalent to the Dolby proposal. Both Baumann et al \cite{2021gpr} and Craig and Singh \cite{craig2011consistent} discuss a consistent histories picture where we are only allowed to calculate two-time probabilities in the case where the two measurements belong to the same consistent history, i.e.  the measurement operators commute when compared at the same instant of time. Finally,  Giovannetti et al propose employing generalized measurements\cite{PhysRevD.92.045033} -  instead of applying the usual measurement operator which collapses the wavefunction into a unique basis state, we apply a unitary transformation which has the effect of recording the result of the measurement in a quantum register without any measurement collapse, so the register itself ends up in a superposition of possible outcomes.  We can then do the same with the second measurement at the later time, thus recording both results in the register, and then subsequently we can perform a measurement on the register to determine the actual pair of outcomes. 
 
 All of these methods reproduce the results predicted by standard quantum mechanics (except possibly in special cases like Wigner's friend experiments, where even standard quantum mechanics does not yield an unambiguous result). But do any of them point us to a viable interpretation of the PW probabilities?  The approach based on relational observables has the appealing feature that it does not require us to add anything to the PW framework, and hence the results of ref \cite{2021trinity} show that at least mathematically speaking Kuchar's objection can be addressed without going outside the original framework. However, because this approach adds nothing to the PW picture it also doesn't offer  any new physical insight which could help solve the interpretational problems discussed in the last section - that is to say, although ref \cite{delahamette2021perspectiveneutral} emphasizes that the relational observables encode gauge-invariant conditional probabilities of the form `the probability that the operator $F$ has some value $f$, given that the frame has orientation $g$,' it remains unclear what those conditional probabilities are probabilities \emph{for} and what the locution `given that' means, since there is no mechanism to select and actualise one particular measurement outcome within this picture.  
 
 Dolby's approach, although formally equivalent, seems a little more physically transparent  in  that it postulates a collapse which could possibly represent a measurement. However, it's unclear whether we're supposed to take this collapse literally, or whether we should  say that the measurement operator plus additional operator are to be understood all in one as some sort of nonphysical transformation; and either way, there does not seem to be any time at which this whole process could take place, given that  the PW formalism presents us with an eternal  picture in which the state of the whole universe is given once and forever.  As argued by Callender and Weingard in regard to the problem of projection operators within quantum cosmology, `\emph{the physical meaning of these projection operators needs to be clarified. Why is the wave function projected, and what selects the associated eigenvalues? ...  If `something else' is involved in the answers, we wish to know what it is. Until then, the purely formal application of projection operators to regain time is not physically justified.}'\cite{CWxx} So although this method does work mathematically, it doesn't offer us any natural way to understand what the probabilities it predicts actually describe.
 
The consistent histories picture seems a little easier to interpret because it allows us to think of the PW probabilities as being probabilities for one course of history or another to occur.  However, this approach prevents us from calculating probabilities at all for a large range of two-time measurements - for example, ref \cite{2021gpr} shows that when applied to Wigner's friend experiments it would allow us to calculate probabilities only for the comparatively uninteresting case where Wigner's measurement is non-disturbing. Moreover, the consistent-histories approach does not yield just \emph{one} set of consistent histories: the formalism  defines a large number of consistent sets of histories, together with a probability distribution across the histories in each set, so even if we select one history from each set according to the prescribed probability distribution, we are left with a large number of mutually exclusive histories\cite{1996consihit}. If we take the `many-histories' view where all of these histories are supposed to occur, then we will presumably find that there are some histories in which some observer measures outcome $O$ when the clock $C$ reads time $t$, and others where the observer measures some other outcome $O'$ when the clock $C$ reads time $t$' and others where no measurement occurs at all at this time, so we will have nothing definite to say about what the observer actually observes, thus undermining the attempt to define any `correlations' between clocks and systems. Moreover, we don't  have amplitudes or probability distributions over consistent sets (probabilities are assigned only \emph{within} sets), so we don't even have access to  the strategies employed in the Everett case to make sense of probabilities under circumstances where all outcomes occur. And finally, it doesn't seem that the consistent histories picture provides us with any representation of observers measuring clocks   -   in order to say anything about the temporal experiences of observers within the universe it will be necessary to include interactions in which observers actually measure the clocks and systems in question, which of course is exactly what the proposal of Giovanetti et al\cite{PhysRevD.92.045033} offers us.

\subsection{The final-measurement interpretation \label{gm}} 

The Giovanetti proposal  requires us to add extra structure to the PW formalism, as we must  include some  interactions in which clock and system states are written to a register. However, from an interpretational standpoint something of this nature would probably have been necessary in any case, as evidently  it cannot be operationally meaningful to define time in terms of correlations between clocks and systems if nobody ever observes the clocks. Indeed, from the interpretational standpoint Giovanetti proposal has an obvious advantage over the other approaches to two-time probabilities, as  it is the only one of these calculation methods  which straightforwardly fulfils the desideratum of providing a representation of observers actually observing or measuring clocks. We reinforce that although the proposal is based on a notion of `generalized measurement' it does not  commit the sin of ascribing a special privileged role to observers or measurements, as other approaches to quantum mechanics have been criticised for doing;  for a `generalized measurement,' unlike the standard notion of quantum measurement,   is simply a feature of the usual unitary quantum dynamics, and thus we are not required to postulate an additional mechanism of physical collapse which occurs only when an observer gets involved. Any physical interaction which maps information in certain degrees of freedom to other degrees of freedom counts as a measurement in this sense, and  thus although observers may get involved by deliberately setting up measuring devices which map information about a quantum system to the `register' represented by their own brain, observers aren't necessary, because natural processes like decoherence are perfectly capable of performing these sorts of mappings without any help from observers. 

That said, there are a couple of technical caveats to be made about the  Giovanetti proposal. First, ref \cite{PhysRevD.92.045033} considers only the case of ideal clocks (for which the time observable is sharp); it is yet to be seen how the approach would work in the context of non-ideal clocks. However, since it seems inevitable that a fully-developed version of the PW formalism will include representations of clocks being observed, we think it is reasonable to assume that some generalisation of this method will also work for non-ideal clocks - if a non-ideal clock couldn't be `read' in this way, we are doubtful that it could be considered a clock at all in any operationally meaningful sense.

Second, although deferring the measurement of the register until after both experiments have been completed ensures that we will get the correct probabilities for that pair of experiments,   if we wish to   also get correct probabilities for any  subsequent experiments conditioned on the results of earlier experiments we will have to defer the collapse process until after those experiments as well, maintaining the register in its superposition state until all the experiments we're interested in have been performed. It follows that if we want this scheme to produce correct joint probability distributions for all measurements and measurement-like events across the whole of history, we will  have to defer the collapse process until after all such events have taken place. 

Third, in the original Giovanetti proposal the conditional probability for a measurement at time $t_2$ to have outcome $n$ given that a measurement at time $t_1$ had outcome $m$ is given by the following expression: 

\[  \frac{ \langle \langle \Psi | t_2 \rangle \langle t_2 |\otimes \Pi^n \otimes \Pi^m | \Psi \rangle \rangle}{ \langle \langle \Psi | t_1 \rangle \langle t_1 | \otimes \Pi^m | \Psi \rangle \rangle } \] 

As noted by ref \cite{2021gpr}, it's somewhat puzzling that this expression involves dividing by the individual modulus for the first measurement at time $t_1$ when this measurement didn't really take place at time $t_1$. Moreover, although this method does indeed reproduce the predictions of standard quantum mechanics in all the cases where standard quantum mechanics makes an unambiguous prediction, ref \cite{2021gpr} shows that if we try to apply it to a Wigner's friend scenario the result of this prescription isn't always a well-defined probability. Ref \cite{2021gpr} observes that we can solve both of these problems by replacing $t_1$ with $t_2$ everywhere in the expression, but then of course it isn't a two-time probability any more. This might look discouraging given that the initial intention here was to define probabilities across events taking place at different times, but in fact it may simply be regarded as an indication that we should take more literally the idea that this equation describes not two measurements performed at different times, but rather a single measurement which resolves both of the results of the individual experiments at once. 

These technical considerations point to an alternative way of thinking about these probabilities: we should simply   defer \emph{all} measurements until the end of time. That is to say, every interaction which has the form of a measurement should be written as a  generalized measurement operator which maps the outcome of the measurement and also the reading on a suitable clock to a register, without actually collapsing the wavefunction to select a unique outcome and time. We then simply evolve the quantum state forwards in time, applying more generalized measurements of this form for every measurement interaction, and writing the results and corresponding times to a blank section of the register. Thus when the register is finally   measured it will  contain a variety of entries of the form `outcome $O$ was observed at time $t$.' The probabilities ascribed by the PW formalism to pairings of measurement outcomes and times can then be interpreted as probabilities for the register to display the corresponding readings at the end of time, and the experiences of conscious agents can be understood as  supervening not on the quantum state but on the actual course of history recorded in the register. We reinforce  there is no need to imagine there exists an observer external to the universe who performs the final measurement: the term `measurement' is simply a way of describing a process in which some particular state of the register is selected and the course of history it records actualised, so the result of the measurement   dictates  the content of reality - in Kent's words, `\emph{an event occurs if and only if it leaves effective records in the final time ... measurement}.'\cite{Kent} Thus the probability distribution over the register is simply a set of probabilities for different courses of history to be actualised, and the universal quantum state is not itself an element of physical reality; it is simply a mathematical tool that we can use to calculate this probability distribution.

There are precedents for this kind of approach in the literature.   Kent \cite{Kent, 2015KentL, 2017} has proposed a `final-measurement' approach to solve the Lorentzian quantum reality problem for relativistic quantum mechanics -  for example  ref \cite{2015KentL} postulates  a final measurement of the density of mass-energy across spacetime, and then stipulates that the actual contents of reality (`the beables') are given by a mass-energy distribution over spacetime such that the value at a given point $x$  is equal to the expectation value of the  stress-energy tensor at $x$, conditional on the outcome of the final measurement outside the future lightcone of $x$. Similarly, ref  \cite{2007mep} reaches a very similar conclusion with respect to the problem of assigning probabilities to sequences of events in general-relativistic quantum mechanics: the authors envision the outcomes of sequences of measurements being recorded in the state of the measuring apparatus, which plays the role of the `register'. It is shown in ref \cite{2007mep} that the probabilities for sequences of measurements thus defined will always be a Dirac observable, and thus these probabilities can legitimately be employed in the context of the PW formalism.  

Formally, this way of thinking about two-time measurements can be represented in the constraint formulation by adding terms of the form $K_{SRC_x} \delta(T_{C_y}- t_m)$\footnote{We note that the use of a delta function in this kind of expression, while common in the literature, is technically problematic: trying to extract relational variables from such a constraint requires us to combine a distribution with a function in a way that is not generally well-defined. Thus it seems likely that in fact the delta functions should be replaced with something like a Gaussian, sharply peaked around the time $t_m$. This is in any case more physically realistic, as the delta function is being used here to indicate that an event occurs when a clock reads a certain value, but since clocks generally cannot be read with infinite precision, any real implementation would necessary have some spread in the exact time of the event. However, since we will not be solving the constraints here we will continue using delta functions for simplicity.} to the constraint, with $K_{SRC_x} $  a generalized measurement operator defined such that: 

 \[ e^{-iK_{SRC_x}} | \psi \rangle | t \rangle_{C_x} | \phi \rangle_R = \sum_i M_i \int_t  | t' \rangle \langle t' |  | \psi \rangle | t \rangle_{C_x} | i \rangle_{R_1} | t' \rangle_{R_2} =   \sum_i M_i  | \psi \rangle | t \rangle_{C_x} | i \rangle_{R_1} | t \rangle_{R_2}  \] 
 
where $| \phi \rangle_R $ describes the blank register prior to this operation, and we split the register $R$ into two  parts $R_1, R_2$ in order to record the result of the measurement on $S$ in $R_1$ and the time according to clock $C_x$ in $R_2$. Here $ \delta(T_{C_y} - t_m)$ is a delta function acting on the time operator $T_{C_y}$ for some clock $C_y$;   the generalized measurement thus occurs at the time when the clock $C_y$ reads $t_m$. We could choose to have $C_x = C_y$, in which case the  time reading written to the register $R_2$ will always be $t_m$, but we could also choose to make $C_x$ and $C_y$ different clocks such that the resulting reading on the register represents information about the time reading on the second clock  $C_x$ relative to clock $C_y$. We then suppose that the states of $R_1$ and $R_2$ are preserved until the end of time so that the final measurement on the register resolves a definite reading $(i, t)$ which represents the measurement outcome $M_i$ being obtained at the time that clock $C_x$ reads time $t$.

Now, there is a complication associated with applying a final-measurement approach in the context of the Page-Wootters formalism. Namely, the whole point of the PW formalism is that there exists no universal time, just a range of different measures of time associated with different possible internal clocks. So if there is no universal time, how can we make sense of the notion of `the end of time'? Won't we find that different choices of internal clock give us a different result for the state of the universe at the end of time? In fact, a problem of this kind arises even if we are dealing with the simpler case of  quantum mechanics in Minkowskian spacetime with some well-defined notion of universal unitary evolution, because different choices of reference frame will give us different final time-slices and hence different results for the final measurement. Kent's solution to this problem is essentially to put the measurement on some time-slice in the distant future and then take a limit as that time-slice   tends to the infinite future\cite{Kent}. This limit will be well-defined provided that at some point in the distant future the universe ceases to undergo any meaningful change or interaction. For example, Kent envisions a scenario in which, due to the expansion of space, eventually all particles and fields become effectively non-interacting. Beyond this point in the history of the universe there are no further interactions, just unitary time evolution, and since unitary time evolution will not alter or destroy any of the information stored in the `register,' it doesn't matter on what time-slice the final measurement takes place as long as it is past the time at which all particles cease to interact. 

In the PW picture, we can't simply put the measurement on some time-slice since different clocks will say different things about the state on that time-slice, but what we can do is take a limit as $t \rightarrow \infty$ for every individual clock in the universe, disregarding any clocks which only exist for a finite amount of time. (Ref \cite{Kent} provides a  technical description of how these limits can be constructed in a relativistically covariant way). We will assume that each of these limits is well-defined and that furthermore they each give rise to the same probability distribution over readings in the `register.'\footnote{Note that in order to adopt this strategy we do not need to assume that all the clocks in the universe have some prima facie orientation which distinguishes between the two possible directions of time, because if we take it that there is some beginning of time, as indicated by our current understanding of cosmology, it follows that the readings on the clocks can go to infinity in only one of the two possible temporal directions, so  the limit  $t \rightarrow \infty$ will necessarily single out the same temporal direction for all the clocks. We do have to assume that the set of all clocks  is at least well enough coordinated that there are two distinguishable time orientations; but clearly in order for the PW approach to be empirically adequate it must  reproduce the dimensionality of spacetime at least at an emergent level, so this assumption is reasonable.} As in Kent's example, we conjecture that this will be the case provided that at some point in history all particles and fields become non-interacting: beyond this point we will get the same probability distribution over register readings  regardless of time, since unitary time evolution will not alter or destroy any of the information stored in the register,  so in the limit as  $t \rightarrow \infty$ all of the possible relational states relative to all of the different clocks in the universe will give rise to the same probability distribution. As noted by Kent\cite{Kent}, there are some indications in current cosmology that eventually all particles and fields will indeed become non-interacting, but we would not go so far as to claim that this has been proven; we regard it as an assumption which we adopt in order to arrive at a transparent operational interpretation of the PW probabilities, or perhaps even as a novel prediction of  this interpretational approach.

This approach offers solutions to the  problems we  identified in connection with defining clock-system correlations in the PW framework.  First, this interpretation allows us to understand the meaning of `given that' in statements like `given that clock A shows time $t$,' because the clock and system are now linked together by the operation which writes the measurement outcome and time onto the register: hence rather than appealing to illegitimate quasi-temporal notions to explain the relationship between states of the clock state and states of the system state, we can make use of pairs of times and measurement outcomes  recorded together in the register.  Second, there is a clear sense in which events can actually be observed, since the operation associated with an event is now a generalized measurement operator which writes the result of a measurement onto a register, and that register might sometimes be the brain of an observer.   Finally, we no longer have to say that every outcome of a measurement always occurs,  as the measurement on the `register' at the end of time selects out a state of the register which represents a single observer observing a single outcome, in accordance with the common-sense notion of `event.' So we don't have to appeal to decision theory or any kind of subjective considerations: we can understand the PW  probabilities as just the usual sorts of probabilities which describe  the outcome of some uncertain event. It is then straightforward to define correlations between clock readings and measurement outcomes based on the  probability distribution over  register readings defined by the universal quantum state. Since the experiences of conscious agents supervene   on the actual course of history which is selected in the final measurement, we now have a clear picture of the way in which the PW probabilities constrain the experiences of agents inside the universe, as they help to determine which course of history is made actual.

Evidently,  interpreting the probabilities in this way does require us to add some additional structure and additional physical hypotheses to the straightforward PW formalism, and this might be regarded as a disadvantage of the approach. However, it seems more or less inevitable that some additional structure must be added if a single-world realist picture is to be extracted from the PW framework, and thus we are willing to accept this sort of additional structure. Obviously we do not claim that this is the only possible interpretation of the PW framework, or even the only possible single-world realist approach - for example, we are also optimistic that there would be a feasible approach based on the framework of relational quantum mechanics. But what we particularly want to reinforce is that  because the final measurement occurs at the `end of time,' the full formal structure of the PW framework is preserved unchanged under this interpretation. Other interpretational approaches do not necessarily have this consequence: for example, the Bohmian approach  advocated by Callender and Weingard\cite{CWxx} and Goldstein and Teufel\cite{Durr2013} requires us to accept a preferred foliation or place additional constraints on the set of allowed wave-functions to counteract the fact that  Bohmian canonical quantisation leads to a spacetime which depends on our choice of lapse function. So there seem to be very good reasons for proponents of the PW framework to adopt the final-measurement interpretation in order to ground the operational significance of their results:  in the final-measurement interpretation, the \emph{interpretation} of some of the results produced by the research programme may be different, but their mathematical correctness is not in doubt, and therefore those who   favour a single-world realist approach can still make sense of the PW framework and the various interesting results that have been obtained within it.

\subsubsection{Comparison to the perspective-neutral collapse} 

The reason that a measurement at `the end of time' makes sense as a way of extracting a single-world realist picture out of the universal quantum state is because,  as discussed by Kent in refs  \cite{Kent,2015KentL},    our brains and all the objects which feature in our macroscopic experience leave a large number of traces   on the environment via the mechanism of decoherence and thus even when we are long dead and all the historical records of our lives have been destroyed, the memories once stored in our brains will still be recorded  in the state of some part of the universe, so in principle these memories can be resolved to a definite state by a final measurement on the relevant `register.' Obviously our use of the term `register' is overly simplistic - there is not literally a book somewhere  recording pairs of times and measurement outcomes - but any physical system can in principle play the role of the register provided that it does not undergo too much disturbance after the interactions in which the relevant record is formed. Therefore the final-measurement approach ensures that  our brains, measuring instruments and  other macroscopic objects will be resolved by the final measurement to have unique, definite histories, as required for a single-world realist approach. 

In particular, the final measurement mechanism provides  a simple and yet  well-defined and physically motivated way of drawing the line between macroscopic features of reality which must be resolved to a definite state in the final measurement and  microscopic features   which need not be resolved: `macroscopic' features are those which are recorded in the final state, and `microscopic' features are those which are not recorded in the final state and thus can participate in quantum phenomena like interference and superposition. Roughly speaking this means that systems which undergo decoherence will be resolved to a definite state while systems which do not undergo decoherence will participate in quantum phenomena\footnote{We reinforce that decoherence alone is not enough to explain the emergence of a single definite macroreality; decoherence makes density matrices diagonal, thus suppressing interference effects, but the measurement mechanism is still needed to select one of the diagonal elements so we get a unique macroscopic history.}. Indeed it would seem that any empirically adequate single-world realist approach must draw this line in roughly the same place, and therefore we conjecture that most other single-world realist approaches  which succeed in retaining the mathematical structure of the PW formalism will ultimately  look quite similar to the final-measurement approach, so the conclusions we draw using the final-measurement approach in the following sections are also likely to be valid in other similar  approaches. 

In this spirit, we noted earlier that one might hope to extract a single-world realist picture from the perspective-neutral approach by collapsing the   quantum state of the universe at the level of the physical Hilbert space. We now observe that the final-measurement approach has obvious structural similarities to the perspective-neutral collapse approach; both strategies postulate a single wavefunction collapse which is required to select the whole course of history in an all-at-once manner. The key difference is that by making use of records, the final-measurement approach arrives at a unifying, asymptotically perspective-neutral object which \emph{directly} encodes histories that we can collapse to, thus sidestepping the problem of fixing the relationships between collapses which occur in different reference frames. As we have noted, it also straightforwardly provides a solution to the measurement problem by answering the question about where to put the line between the microscopic and the macroscopic. So at least for the moment the final-measurement approach seems like a more complete answer to the intepretational problems of the PW framework than the perspective-neutral collapse.  That said, the two approaches are not necessarily in competition; for proponents of a perspective-neutral collapse, the final-measurement interpretation could be regarded as a heuristic tool which we may employ in the short term to get some idea of what the perspective-neutral collapse might look like, since we do not yet  yet have a formal framework showing how to implement that collapse. 

We discuss a few more interpretative points in appendix \ref{fm}.

\subsubsection{Example: Temporal Nonlocality \label{localisation}}

This is all good news for the PW programme and more broadly the IQRF research programme. First off,  the final-measurement interpretation goes some way towards vindicating the claim that the PW formalism provides an adequate account of time in the  block universe, for it provides us with  a clear operational connection between the PW relative states and the temporal experiences of actual observers. Moreover, we also see that the PW account of time does not in fact depend on the adoption of the Everett interpretation or some kind of non-realist view - those who prefer a `single-world realist' account which allows for unique, intersubjective measurement outcomes also have a way of making sense of the PW formalism. We will now verify this claim by examining the interpretation of a specific result of the IQRF framework within the final-measurement interpretation.

Ref \cite{2020qctloe} discusses  `temporal non-locality' in the IQRF approach. For example, consider the case where an event $E$ occurs in the reference frame associated with a clock $C_B$, i.e. the event $E$ is fixed to occur when clock $C_B$ reads $t$. Suppose now that clock $C_B$ is in a delocalised state relative to another clock $C_A$ -  i.e. the state of  clock $C_B$ relative to clock $C_A$ represents it as being in a superposition of time readings. Ref \cite{2020qctloe} demonstrates that   an event $E$ will therefore be delocalised in time relative to clock $C_A$. But what does that actually mean? Is it just a mathematical statement about the relation between $E$ and $C_A$ or does it have operational consequences for  real temporal experiences? 

We can answer these questions by appeal to the final-measurement  interpretation. In order to represent the event $E$ in the reference frame of clock $A$, we will use a generalised  measurement which is scheduled to occur when clock $B$ reads $t$. This can be expressed using the constraint equation $(H_A + H_B + \delta(t -T_B) K_{R C_A}) | \psi \rangle = 0$, which encodes the fact that when the clock $C_B$ reads $t$ the operator $K_{R C_A}$ copies the time from clock $C_A$ over to the register. Here $T_B$ is the time operator associated with clock $C_B$ as defined in section \ref{intro}; in this case we are only interested in the time of the event and not in any additional measurement result, so we don't need the operator $K$ to also copy over a measurement outcome as defined in section \ref{gm}.   Let   $| \psi_A(0) \rangle_{\lnot A}$ be the state of all systems except clock $A$ conditioned on $A$ reading $t = 0$; then as in ref \cite{2020qctloe}, $| \psi_A(0) \rangle_{\lnot A}$  can be written as $\int d t_B \phi(t_B) |t_B \rangle \otimes | \chi \rangle_{R}$, where the normalised function $\phi$ encodes the localisation of clock $C_B$ relative to clock $C_A$: if it's a delta function then the two clocks  have sharp states relative to one another, whereas if it isn't  a delta function then there is no definite fact about what clock $B$ reads when $A$ reads $0$.
    
Now let us suppose that experiments of this kind are carried out many times and in each experiment the clocks $C_A, C_B$ are reset to some appropriate initial state at the start, so we can expect to see numerically consistent time readings across these different experiments. Then if $\phi$ is a delta function, the generalized measurement operator $K_{RC_A}$ mapping the reading of $C_A$ to the register will always occur when clock $C_A$ reads some fixed value $t'$ and  therefore when the register is measured we will find that all of the registers for the various copies of this experiment read $t'$. On the other hand, if $\phi$ is not a delta function then the state of the register $R_1$ after the generalized measurement operator $K_{RC_A}$ has been applied will be some superposition of different time readings, and and therefore when the register is measured we will find that the records for the various copies of this experiment contain a variety of different time readings.   So the final-measurement interpretation wholly vindicates the claims of the IQRF framework in this instance and indeed puts those claims on a   more solid operational framework: what it is for one event to be delocalised in the reference frame of a given clock is for the time of the event  to exhibit some spread across an ensemble of identical experiments as recorded in the register at the end of time, rather than always showing the same time.

\section{Quantum Reference Frames \label{qrf}}

We will now consider how the foundational and conceptual claims that have been associated with the PW and IQRF frameworks may be affected by a choice of interpretation for those frameworks. We will begin by evaluating  the claim that the IQRF research programme can be understood as a way of thinking about reference frames in the quantum world. As we have seen,  the results of the PW and IQRF research programmes are often expressed using language like  `\emph{the time reference frame of a clock}' and  `\emph{In this situation, how would another clock, say A, describe the time evolution of a quantum experiment?}'\cite{2020qctloe} and the process of selecting a quantum references frame is often described as `\emph{jumping into the perspective}' of a quantum system\cite{2020acop}.  Prima facie this language may seem a little odd, as clocks and individual quantum particles do not literally have a perspective on the world and they do not literally provide descriptions of experiments. So this language must  be understood as shorthand for something else - but what?

The IQRF idea that reference frames are simply individual physical systems is motivated by Einstein's treatment of reference frames, where we are encouraged to think about the operational procedures   used to establish the values of physical quantities in different reference frames using physical objects like rods and clocks. For example, ref \cite{giacomini2021quantum} suggests that we should `\emph{take the view, already advocated by Einstein, that coordinate systems correspond to physical rods and clocks.}' But note that Einstein does not tell us that rods and clocks \emph{are} reference frames - rather a number of rods and clocks are used jointly to operationally establish a reference frame\cite{Einstein1905}.   An alternative way of thinking about reference frames is suggested by the fact that rods and clocks are not just any old physical objects - they are \emph{measuring instruments} which take part in operational procedures that are used to characterise physical events relative to the reference frames. This feature of rods and clocks is obviously crucial to Einstein's use of them, as the whole point of a reference frame is to serve as a background against which we can describe physics, and typically a `description of physics' involves characterising the results of some actual or possible observations. If a description of physics `relative to' some reference frame is not anchored to actual or possible observations then there will inevitably be a worry that it is merely a mathematical abstraction which does not represent anything in physical reality - and  this worry is particularly pressing when the physics we are trying to describe is quantum-mechanical, since there remains a great deal of uncertainty about the ontology of the theory and therefore it's by no means certain that descriptions of the evolution of a wavefunction have any physical meaning outside of the role they play in characterising possible observations.

Of course, in classical physics the measurements can largely be left implicit when we characterise a reference frame, because classical measurements can be regarded as straightforwardly probing classical reality without disturbing it in any way. But this is no longer the case when we are dealing with quantum physics, because measurements and records can't be taken for granted in the quantum world   - performing quantum measurements will generally disturb the evolution of the relevant systems, and no-cloning constraints place limits on the degree to which records can be made\cite{Scaranicloning}, so in a quantum setting it is not necessarily the case that abstract statements about the evolution of one system relative to another system can be converted in any straightforward way to operational statements about the results of measurements.  Therefore  it is important to explicitly include measurements and records in a theory of quantum reference frames, because otherwise we risk ending up with  descriptions which can't possibly correspond to any actual observations. 

These considerations seem to indicate that in order for a IQRF to play the same kind of role as a classical reference frame, we must be able to make sense of the idea that it could have   measuring instruments  which are `carried along' with it, like the co-moving rods and clocks employed in the construction of an idealised classical reference frame. Describing the physics inside an IQRF would then amount to saying what the measuring instruments associated with the reference frame would read under various circumstances (if there were in fact such measuring instruments present). This would allow us to make physical sense of  the notion of `the evolution of some system $S$ relative to the reference frame $F$,' since it could be cashed out in terms of operational procedures, such that if we are in reference frame $F$ and we perform and record a sequence of measurements on the system $S$, the recorded results will be consistent with the specified evolution law.  

We will henceforth take it that the distinctive feature of IQRFs as compared to classical references frames is that they can be in superpositions relative to one another: thus if IQRFs are in general to be operationally accessible we will have to postulate that measuring instruments or measurements or the physical systems recording the results of those measurements can in some sense be in superposition relative to one another. For example, if a particle   is in a superposition relative to a laboratory, the outcomes of experiments performed using the laboratory instruments will appear to be in a superposition relative to the particle reference frame, and conversely if we could somehow manage to associate some measuring instruments with the particle, then the outcomes of experiments performed using those instruments would appear to be in a superposition relative to the laboratory reference frame. 

One might be tempted to suggest that instead of having a distinct set of  measuring instruments for each frame, as with  co-moving rods and clocks in general relativity, we could employ just a single measuring device such that when we transform to a particular IQRF, we also transform the state and observables of the measuring device. So for example, in the particle-laboratory case, rather than associating measuring instruments with the particle itself, we could simply characterize how the outcomes obtained from the laboratory instruments will look in the particle IQRF - i.e. they will be in superpositions.  However, it seems quite difficult to  arrive at an operational understanding of such measurement results, as we ourselves do not ever find measuring instruments or records of measurement outcomes in superpositions and indeed we do not even seem to have a coherent notion of what it would be like to find a measuring instrument or record in a superposition. Moreover, as we will shortly see, the IQRF  research programme tells us that within each reference frame the metric is locally Minkowskian ( i.e. the metric is flat and it is not in a superposition relative to the reference frame), and if this statement is to be understood operationally, then it must be the case that within each reference frame we can perform local measurements probing the metric  which produce results that are not in a superposition relative to that reference frame. So it is reasonable to conclude that  if IQRFs are in general to be operationally accessible, we will sometimes have to postulate cases where two distinct measuring instruments and the physical records of their results are in superposition relative to one another.

We note that our emphasis on the operational features of reference frames is not predicated on the philosophical view sometimes known as `operationalism' which holds that a concept cannot be meaningful unless we have a method of measurement for it\cite{sep-operationalism}. It is not our intention to claim that `quantum reference frames' which are not associated with  measurements are in virtue of that fact not \emph{meaningful} - we merely contend that such constructions are not well suited to play the role of \emph{reference frames}, because the function of reference frames in our physics is intimately tied to operations of observation and measurement, so   a theory of reference frames must ultimately say something about  measurements and observations. Moreover, we recall that the original motivation for the PW  framework was to define time in operational terms through measurements of quantum clocks, so if the framework is to fulfil this promise, the reference frames that it employs must be given some operational interpretation.

\subsection{The Quantum Equivalence Principle \label{QEP}} 

Against these arguments, one might object that as noted in section \ref{intro}, choosing a reference frame can be regarded as a choice of gauge-fixing procedure, and for other types of gauge-fixing procedures we don't typically insist that a gauge must have operational significance in terms of actual or possible measurements. Thus for example ref \cite{2021qrft} argues that matters of this kind can be left until later: \emph{`One may finally ask whether observers who are part of the theory may in fact have good operational access to that chosen frame of reference, but this is an additional (though important) question that we here regard as secondary.'}  And of course we have no objection to the IQRF methodology if it is merely regarding as gauge-fixing. There are certainly good reasons to wish to have access to these sorts of alternative descriptions: for example, ref \cite{hoehn2021internal} notes that this may help for certain sorts of calculations: `\emph{if we do not know how to handle the quantum case, let us perform a IQRF transformation that renders some of the subsystems classical.}'  However, the use of the terminology `reference frame' for this gauge-fixing procedure is not entirely innocuous, because it gives the impression that `quantum reference frames' play a similar role in the theory to classical reference frames, and thus it it risks giving undue weight to descriptions of the world that are not in fact operationally or physically meaningful. In particular, reference frames play a key role in certain formulations of Einstein's equivalence principle, and therefore proponents of the IQRF reference programme have drawn connections between their results and the equivalence principle: for example, ref \cite{giacomini2021quantum} argues that   `\emph{there is no fundamental incompatibility between the principles of (quantum mechanics and general relativity), provided that we revise the formulation of the Equivalence Principle in such a way that it holds in all reference frames that quantum theory allows for.}' Given the well-known difficulties associated with the attempt to unify quantum mechanics and general relativity, this is a very strong claim, and one might worry that to some degree the proposed solution relies on equivocation around what a reference frame is. 

The `quantum equivalence principle' proposed by refs\cite{giacomini2021quantum, giacomini2021einsteins} states that `\emph{In any and every Quantum Locally Inertial Frame, anywhere and anytime in the universe, all the (nongravitational) laws of physics must take on their familiar non-relativistic form.}' Here, a Quantum Locally Intertial Frame is obtained by implementing a unitary IQRF transformation such that the gravitational field, and the superposition of gravitational fields, is  locally Minkowskian in the quantum reference frame associated to the position of some quantum particle. That is to say, a QLIF is defined with respect to an individual physical system rather than in terms of actual or possible measurement operations.  On the other hand,  the `reference frames' employed in general relativity typically \emph{are} defined in terms of actual or possible measurements, so some further work is needed before we can verify that the quantum equivalence principle  is suited to play the same kind of role in quantum gravity as the ordinary equivalence principle plays in general relativity.

At this juncture we should acknowledge that even if we confine our attention to General Relativity, there are a number of different ideas going by the name of `equivalence principle,' - ref \cite{2010dqmcw} demonstrates that there are at least four different classes of equivalence principle. However, it seems clear that the version of the equivalence principle  invoked by the `quantum equivalence principle' above  is what ref \cite{2010dqmcw} describes  as the `small enough regions' formulation: i.e. the notion that in any general relativistic spacetime, the metric in a sufficiently small region of a given point will look Minkowskian. Thus we will henceforth focus on this specific version of the equivalence principle, which is sometimes known as the `strong equivalence principle' (SEP)\cite{pittphilsci17709}. What, then, is the function of the SEP in general relativity? \footnote{It's worth nothing that there has been considerable debate about whether the SEP is universally true - it has been pointed out that tidal effects may spoil the Lorentz invariance unless the `region' is just a point\cite{synge1960relativity}, and more recently it has been noted that minimal coupling may also lead to apparent violations of the equivalence principle\cite{pittphilsci14589}; the fact that the SEP can apparently be violated even in the classical regime casts some doubt on the idea that we should expect it to be universally true in the quantum regime.}

In Einstein's own writings, the SEP is understood in operational terms as a description of possible measurement procedures: he famously characterised the equivalence principle in terms of what an  observer would experience: \emph{`for an observer in free-fall from the roof of a house there is during the fall—at least in his immediate vicinity—no gravitational field.'}\cite{Einsteinmethods} Expanding on the important role of these operational considerations in the theory, he wrote in 1921, `\emph{According to the special theory of relativity the coordinates $x, y, z, t$ are directly measurable via clocks at rest with respect to the coordinate system. Thus, the invariant $ds$, which is defined via the equation $ds^2 = dt^2 - dx^2 - dy^2 - dz^2$, likewise corresponds to a measurement result. The general theory of relativity rests entirely on the premise that each infinitesimal line element of the spacetime manifold physically behaves like the four-dimensional manifold of the special theory of relativity.}'\cite{Einsteinletter} This line of thought has been taken up in particular by Brown\cite{Brownrelativity}, who argues  that  the crucial function of the strong equivalence principle in General Relativity is to make a link between the metric and the behaviour of actual clocks and rods: the metric \emph{`attains its chronogeometric significance in GR ...  by way of the strong equivalence principle.'}\cite{pittphilsci14589} Similarly Knox writes that the SEP \emph{`expresses just that fact about our matter theories that must be true if systems formed from appropriate matter are to reflect the structure of the metric field, that is, if phenomenological geometry is to reflect the geometry of the metric field,'} and further that \emph{`the whole point of the SEP is that it's meant to tell you what it means for experiments to pick out a connection.'\cite{Knox2013-KNOESG}} That is to say, the SEP is important because without it the mathematical framework of GR can't be understood as describing the behaviour of actual rods and clocks, which would largely prevent it from having meaningful observational consequences.\footnote{We reinforce that although it is particularly common for proponents of the dynamical perspective on relativity\cite{Brownrelativity} to emphasize the role of the SEP in defining chronogeometric behaviour,  this understanding of the SEP isn't predicated on acceptance of the dynamical approach: the SEP plays a very similar operational role in the competing geometric approach\cite{maudlin2012philosophy}. If we adopt the dynamical perspective, we would say that the equivalence principle is true because the dynamics of matter fields locally exhibit the same symmetries as the metric; whereas if we adopt the geometric perspective, we would say that the equivalence principle is a consequence of the fact that the metric constrains the matter fields to be invariant with respect to its own local symmetries: either way, it is clear that the equivalence principle is important in large part because it encodes a practical connection between the metric and the dynamics of matter. The two approaches differ on the direction of explanation, but agree that the SEP is significant because of this practical connection.}

 So it seems that the role played by the SEP in general relativity, or at least one particularly crucial role that it plays, is founded on the fact that the `reference frames' to which it refers have an operational significance in terms of actual or possible measurements which can be used to survey the metric. These operational features are not incidental - they are at the heart of the equivalence principle's relevance and importance. Therefore if a `quantum equivalence principle' is to play a similar role in a theory of quantum gravity, it should likewise be defined with respect to reference frames which can be understood in terms of actual or possible measurements which give us operational access to the spacetime metric. 
 
We reinforce that the IQRF research programme is still under very active development, so although the connection between the IQRF formalism and actual or possible observations is not currently very clear, this is a topic which researchers are still working on, and we are confident that interesting answers will emerge in due course.  It is not our intention to anticipate  these  conclusions, so we will limit ourselves to making some general   observations about the way in which the status of the quantum equivalence principle will necessarily depend on a choice of interpretation of quantum mechanics. The crucial point is that  different interpretations of quantum mechanics   differ in  what they say about the possibility of performing distinct measurements within different IQRFs. 

\paragraph{Collapse approaches} 

In the case of interpretations which involve    wavefunction collapses  there is a very concrete sense in which it is not possible for two reference frames each containing its own set of measurements to be in a superposition relative to one another, because measuring  instruments are macroscopic objects and hence  they are, according to these approaches, not able to be in superpositions.  In particular,  the GRW collapse and gravitational collapse interpretations predict  that superpositions inevitably collapse when the objects involved in them get too large. That is to say, because these approaches define a set of beables which by stipulation are in a definite state, they necessarily single out a set of preferred quantum reference frames relative to which the beables are not in superpositions, and from that point of view quantum descriptions relativized to any other frame of reference are strictly false. Thus these approaches tell us that many of the reference frames that we might extract from the perspective-neutral universal wavefunction don't correspond to anything physically real or meaningful. Moreover by construction it is guaranteed that in these approaches macroscopic objects will never be in superpositions relative to the preferred reference frames, and therefore collapse approaches will never allow us to have two distinct macroscopic measuring instruments or observers in superposition relative to one another, so we will never be able to perform distinct measurements surveying distinct spacetime metrics in IQRFs which are in superposition relative to one another. Therefore it would seem that collapse approaches   don't admit the possibility of  operationally meaningful quantum reference frames (except insofar as they coincide with classical reference frames).

\paragraph{No-collapse single-world realist approaches}

The final-measurement interpretation, on the other hand, tells us that the universal wavefunction may indeed contain representations of measuring instruments and observers which are in superposition relative to one another. Thus  all of the  different reference frames that can be extracted from the perspective-neutral universal wavefunction are in some sense physically meaningful, and we can understand transformations between quantum reference frames as gauge transformations relating different possible ways of describing the universal wavefunction.  On the other hand,   the  final-measurement interpretation tells us that when we  have two reference frames superposed relative to one another and we attempt to perform measurements surveying the metric individually in  each of these reference frames, if it is the case that the metrics are macroscopically different then the  records of the two sets of measurements will  be orthogonal, meaning that the final measurement will necessarily reveal a set of register readings which actualises only one set of measurement outcomes. Therefore the final-measurement approach tells us that superpositions of reference frames relative to one another are a real feature of the universal wavefunction, but  nonetheless only a subset of these quantum reference frames will   have concrete operational significance in terms of actual measurements, and in particular we will never find that two reference frames in a superposition relative to one another both come to have direct operational significance.  

Similarly, in the de Broglie-Bohm interpretation there are no wavefunction collapses and hence in a de Broglie-Bohm version of the IQRF framework it follows that all of the different reference frames that can be extracted from the perspective-neutral universal wavefunction are in some sense physically meaningful. However,  the de Broglie-Bohm approach tells us that  measuring instruments and measurement outcomes supervene on the arrangement of the de Broglie-Bohm variables (e.g. positions of particles in the simplest version of de Broglie-Bohm), and since the de Broglie-Bohm variables cannot be in superpositions relative to one another,  once again we will never find that two reference frames in a superposition relative to one another both have direct operational significance.  

The distinction between reference frames which have concrete operational significance and reference frames which do not is important, because   in the case of classical  relativity, one of the main motivations for studying transformations between reference frames is the idea that all reference frames are equally valid:   no reference frame is more `real' than any other reference frame, and therefore we expect that observers should have roughly similar kinds of experiences in different  reference frames. But this is not necessarily the case when we are dealing with IQRFs in a single-world realist approach. For example, the de Broglie-Bohm interpretation and the final-measurement interpretation both tell us that when a particle is superposed relative to a laboratory, \emph{only} the reference frame associated with the laboratory is genuinely an element of physical reality: a reference frame defined with respect to a superposed quantum particle is not `real' in the same way as ordinary physical objects, it is simply an artefact of some mathematical structure that plays a role in defining the behaviour of the physical world. So adherents of these sorts of single-world realist approaches might reasonably feel that  physics should be described relative to the real reference frames composed of physical objects, rather than relative to `virtual' reference frames defined with respect to superposed particles which are not elements of ordinary physical reality. And thus adherents of single-world approaches might not necessarily think it is correct to expect that real reference frames should be `equivalent' to these virtual reference frames in the way defined by the equivalence principle. 

\paragraph{Everett} 
 
In the Everett interpretation, on the other hand,  in principle there could be IQRFs in superposition relative to one another such that each IQRF is operationally accessible to a distinct set of observers inside it. For example, we could have a  `Wigner's friend' type scenario where an observer performs a measurement on a superposed quantum particle and thus aligns herself with the reference frame of the particle, such that both the observer and her measurement results are in a superposition relative to the reference frame of some other external observer. However, recall that decoherence has the effect of ensuring that states involving macroscopic objects very rapidly become diagonal, so even in an Everettian world it would be extremely difficult to actually implement  a scenario where we have two  IQRFs associated with macroscopic objects in superposition relative to one another for any nontrivial length of time. 

Moreover, even if we did manage to implement such a scenario, recall that it is crucial to the empirical success of the Everett view that due to decoherence, observers within one branch of a superposition will have exactly the same experiences as they would have had if the wavefunction had in fact collapsed and there were no other branches, so it is unclear to what extent those observers will need a special quantum version of the equivalence principle to describe their experiences.  For example, consider the QEP as  formulated in ref \cite{cepollaro2021quantum}: \emph{`The local effects of (quantum) motion in a superposition of uniform gravitational fields are indistinguishable from those of an observer in flat spacetime that undergoes a quantum superposition of accelerations.'} In order to say that these two experiences are `indistinguishable' we must first say something about the subjective experience of the observer in both cases: so what does it feel like to experience `a quantum superposition of accelerations?' The Everett interpretation tells us that `an observer that undergoes a quantum superposition of accelerations' will in fact become a set of distinct observers within distinct branches of the wavefunction, each experiencing a different single-valued acceleration, so it would seem that in the Everett picture there is no difference between the subjective experience of `undergoing a quantum superposition of accelerations' and the subjective experience of undergoing a single acceleration. Therefore it's unclear  what it could mean in the Everett picture for the experience of undergoing a superposition of accelerations to be indistinguishable from the local effects of (quantum) motion in a superposition of uniform gravitational fields, unless this is simply to be cashed out in terms of  a mapping between the branches in the two cases, such that each observer undergoing motion in some branch of a superposition of uniform gravitational fields has experiences indistinguishable from those of some  observer in a branch of a quantum superposition of accelerations in flat spacetime. But that is just the same as saying  that `the local effects of  motion in a uniform gravitational field are indistinguishable from the effects of acceleration in flat spacetime,' which already follows from the classical SEP. There don't seem to be any distinctive experiences associated with being in a IQRF which set IQRFs apart in \emph{operational terms} from classical reference frames, and hence there is no extra work for an operationally construed quantum equivalence principle to do here.

Given that it is decoherence which appears to block Everettian observers from having subjective experiences whilst in superpositions which differ from the experiences they have when not in superpositions, one might be inclined to suggest that the operational significance of the quantum equivalence principle could be understood in a counterfactual sense: in some theory like Everettian quantum mechanics in which observers were \emph{not} subject to decoherence, the experience of being in a superposition of accelerations would be indistinguishable from the experience of being in a superposition of uniform gravitational fields. But it's very hard to make sense of this because, as often observed in the literature on interpretations of quantum mechanics, it's very hard to even imagine what the subjective experience of `being in a superposition of accelerations' would feel like\cite{albert1994quantum}: the consensus seems to be that the only way we can make sense of this notion is to suppose that each branch is associated with a distinct conscious experience. So it's unclear who the subject of experience could possibly be in the `indistinguishabilility' claim made in ref  \cite{cepollaro2021quantum}, even if we allow ourselves to consider an alternative decoherence-free quantum world. (Indeed, it seems unlikely that a decoherence-free quantum world would contain any conscious observers at all!)

\subsection{Summary} 

Thus we see roughly three possible cases for the operational status of quantum reference frames: 1) reference frames can't be in superpositions relative to one another at all; 2) the universal quantum state may contain representations of reference frames which are in superposition relative to one another, but   in such cases  only one of the reference frames can be operationally accessible; and 3) operationally accessible reference frames can be in superposition relative to one another, but  from a subjective point of view being in a IQRF is just the same as being in a classical reference frame, so in most cases the classical SEP will already be enough to characterise the experiences of observers in such reference frames. Evidently the status of the `quantum equivalence principle' will be different in each of these cases: however, in none of these cases does the quantum equivalence principle appear to play the same kind of role as the classical SEP.  (In appendix \ref{Penrose} we discuss a particular case where the difference between the classical SEP and quantum equivalence principle becomes important.)

We reinforce that the points made in this section are not intended as a criticism of the IQRF framework itself. The basic reasoning that we should be able to describe physics in a way that does not depend on the existence of a hypothetical reference frame outside the universe seems impeccable, and  the IQRF  formalism is doing important work to show how this can be achieved using a gauge-fixing process to deal with the redundancy arising from the absence of a universal frame. However, it seems that in many cases it is not really viable to understand this form of gauge-fixing in  operational  terms, so it is not particularly helpful to think of the resulting description as being relativized to a `reference frame' in the usual operational sense. Not every meaningful statement about physics needs to be characterised directly in operational terms: it is enough that we can see how relevant operational features would appear in some appropriate limit. 

Indeed, instead of thinking of the IQRF framework as characterising reference frames directly, we can potentially regard the formalism as a route to showing how  `reference frames' and other operational features emerge from the perspective-neutral view in the limit as the size of the systems involved becomes large. In this limit, and \emph{only} in this limit, we can think of the gauge-fixing procedure as    `jumping into the perspective' of an internal subsystem, because if a system is macroscopic we can ascribe to it a `perspective' which corresponds to an operationally meaningful description. This suggests that rather than postulating a distinctive `quantum equivalence principle,' the IQRF framework should perhaps instead be seeking to show  how  the ordinary classical equivalence principle emerges in the appropriate limit. In this connection we recall Knox's verdict on the SEP: in ref  \cite{KNOX2013346} she argues that the SEP has to be understood as `\emph{contextual and approximate}'  and that `\emph{phenomenological geometry is exactly the kind of thing that could be emergent},' which leads her to question whether it makes sense to think of the SEP as holding in the context of quantum gravity, given that quantum gravitational theories are likely to involve a characterisation of reality somewhere below the level at which phenomenological geometry emerges. The conclusions we have reached here on the `quantum equivalence principle' seem to uphold Knox's arguments: we have seen that   `reference frames' themselves are necessarily an emergent phenomenon, and thus any equivalence principle making essential use of reference frames must also be emergent. 

Poulin demonstrates this kind of emergence in ref \cite{2006poulin}, which offers a description of a `measurement' in relational terms: a quantum particle is coupled to a `gyroscope' and then we check whether the spins of the system and gyroscope are parallel or antiparallel, averaging over all possible directions to eliminate non-relational facts. Poulin shows that in the limit as the gyroscope becomes very large, the probabilities for the two outcomes just reduce to the amplitude squared for each of the two possible spin directions in the particle's non-relational quantum state, i.e. in this limit we arrive at the standard quantum description.  Poulin's work is done in the context of the quantum information approach to reference frames (see appendix \ref{others}) where we use statistical averaging rather than coherent averaging as in the IQRF framework, but nonetheless the general principle that as a system becomes macroscopic the physics relative to it approaches the standard non-relational quantum description is presumably also applicable in the IQRF approach, and this provides an important constraint on the operational credentials of that framework.  Any `operational' approach to physics must take account of the fact that  the way in which we ourselves are embedded in reality places certain limits on the operations we can perform; in this case   the key point is that we are macroscopic objects who are always subject to decoherence effects, and therefore we will never be able to directly observe relational effects which hold only when all the systems involved are microscopic. The act of observation necessarily introduces a macroscopic reference frame, and as shown by Poulin, that macroscopic reference frame introduces qualitatively different features into the relational description, which inevitably makes it difficult to determine concrete operational features of quantum reference frames which differ from the operational features of classical reference frames.

\subsubsection{Less operational alternatives}

Given the considerations discussed in the previous section, potentially the gauge-fixing procedure of the IQRF framework would be better understood as characterising   `quantum coordinate systems,'  rather than reference frames, since a coordinate system is not required to have any particular operational significance. Then rather than a quantum generalisation of the SEP we should perhaps be aiming for  a quantum generalisation of the principle of general covariance, which would presumably say something like `The equations describing the laws of physics should have the same form in all (quantum) coordinate systems, or IQRFs,' - for example, in ref  \cite{hardy2019implementation} Hardy embarks on the project of writing the laws of quantum mechanics in a  tensorial form and expressing them in  a bundle of four-dimensional Riemannian spacetimes, so a principle of general covariance can be applied. However we reinforce that  as noted by Kretschmann\cite{Krestschmanngc}, the principle of general covariance is in some ways   trivial - it is possible to write virtually any classical spacetime theory in such a way that it `has the same form in all coordinate systems,' since this can be guaranteed by writing the laws in a tensorial form and expressing them in a four-dimensional Riemannian spacetime. The SEP, on the other hand, certainly has non-trivial content: as noted by ref  \cite{Brownrelativity} the SEP does not even follow from the Einstein Equations of general relativity, it is a further assumption needed to give operational meaning to the theory. So the equivalence principle and the principle of general covariance  play distinctly different roles in general relativity, and thus it is important to distinguish clearly between a quantum principle of general covariance and a quantum equivalence principle, which would presumably have quite different roles to play in quantum gravity.  

If we do still want to postulate a  specific quantum equivalence principle, we might perhaps  move to  a `quantum equivalence principle' defined using analogues of measuring instruments which are not  macroscopic. For example, although ref \cite{cepollaro2021quantum} initially describes the three aspects of the quantum equivalence principle in terms of the experiences of (presumably macroscopic) observers, the way in which it is proposed that we test the quantum equivalence principle is by employing  superpositions involving  particles with an internal degree of freedom which can be treated as a clock. It is then argued that we will have a violation of the quantum equivalence principle if \emph{`in the QLIF of system A, the clock behaves differently according to where it is placed in the gravitational field.'} 

Now,  in order to use the internal degree of freedom as a measuring instrument we must make some observation of it, and obviously any single-world realist interpretation will insist that we always observe exactly one out of the set of possible superposed clock readings, so such interpretations must still insist that in this case it is impossible to actually perform distinct sets of measurements surveying different metrics in the different IQRFs. However,  instead of observing the clocks directly, we might consider performing some kind of interference experiment between the branches of the clock and then observing the result of that experiment. This is exactly what  ref \cite{cepollaro2021quantum} proposes, and it is shown that various kinds of violations of the quantum equivalence principle can be expected to have characteristic effects on the results of the interference experiment. So it seems that according to   ref \cite{cepollaro2021quantum} the quantum equivalence principle does not in fact have direct operational significance, but rather it is to   be accessed indirectly by performing measurements which   give us information about what the relationship between the measurement results \emph{would have been} if we could have performed distinct measurements for each of the frames. The operational meaning of the quantum equivalence principle then reduces to the claim that the results of this interference experiment will be consistent with what we would expect if  the internal degrees of freedom of the particles  each individually surveyed the metric inside their own IQRF in the same way that clocks are expected to survey the metric in ordinary classical GR.  

Note that it seems crucial to this formulation of the quantum equivalence principle that the IQRFs in question are not accessed by direct measurement - for example, we don't actually perform readings in the QLIF of system A to see if the clock behaves differently according to where it is placed in the gravitational field. For if we did that, then the ordinary classical SEP would come into force, mandating that the clock must not behave differently according to where it is placed in the gravitational field. Thus if the quantum equivalence principle is to have any content over and above the classical SEP, it must be understood as characterising readings on measuring devices at times when they are not actually being read. This immediately indicates that a quantum equivalence principle formulated in these terms cannot play the same kind of role as the SEP, which as we have seen is particularly important to GR because of the connection it makes between \emph{actual} observations and the underlying spacetime metric. Moreover, this way of thinking about IQRFs takes an unusually robust view of the properties of systems at times when they are not being measured. For recall that the contextuality theorems in quantum mechanics\cite{KochenSpecker, Spekkens} have demonstrated that it is not always possible to think of quantum systems as having definite well-defined properties at times in between measurements: in view of this  feature of quantum mechanics, the claim that we can meaningfully think of an individual quantum particle as having a well-defined `perspective' or `reference frame' and a well-defined reading on its internal clock at times in between measurements is a highly nontrivial interpretational assumption. And if we don't  think of particles   as having well-defined states, properties or perspectives at all times, then it seems we must take a somewhat deflationary view of the quantum equivalence principle - i.e. we must regard it as a heuristic principle which guides the predictions we make for actual measurements, rather than as a literal description of physics within inaccessible reference frames.

\subsection{Example: Temporal nonlocality again}

Returning to the case of the final-measurement interpretation, let us consider what would be required in that picture in order for a IQRF to correspond to a physically and operationally meaningful reference frame. At the very least, the IQRF must be equipped with some kind of register,  such that we can perform generalized measurements as discussed in section \ref{gm} which map the results of measurements on the systems to records on the register.  So for example in the case where the reference frame is a quantum clock $C$, we should add   a quantum register $R$ to our description, and then measurements performed on some system $S$ in order to describe its evolution relative to the reference frame associated with clock $C$ can be represented as generalized measurements which are added to the Hamiltonian constraint as terms of the form $\delta(T_C - t) K_{SC}$, where as in section \ref{gm} the   operator $K_{SC}$ copies the reading on the clock $C$ plus the result of some measurement on the system $S$ onto the register $R$. Such operations may be carried out at  several different times throughout history, so the full history state represents the register associated with clock $C$ as being in a superposition of various different histories, and what is written on the register when it is measured at the end of time can be regarded as  the observed time evolution of system $S$ relative to clock $C$. Talk about the `evolution of system $S$ relative to frame $F$' then has a transparent operational interpretation: it can be understood in terms of statements about  what the register would be expected to read if appropriate generalized measurements were performed and the register were then measured at the end of time.  Whether or not those measurements are actually made, including the register explicitly in our description ensures that descriptions of physics in the relevant reference frame have some physical content.

One may then ask whether the altered definition of a `reference frame' suggested by the final-measurement interpretation substantially changes any of the results of the IQRF research programme.  For example, consider the time localisation scenario discussed in section \ref{localisation}. There we discussed the case where only clock $C_A$ is associated with a reference frame, but ref \cite{2020qctloe} also discusses a putative case where we have one reference frame associated with each of the two clocks $C_A$ and $C_B$, and each reference frame is delocalised relative to the other. For example, this can be done by having  the first clock $C_A$ be an atomic clock  and the second clock $C_B$ be a clock associated with the laboratory, where $C_A$ is entangled with some other atomic clock from the perspective of the laboratory and the two atomic clocks are sent through an interferometer in a gravitational field, thus providing what ref \cite{cepollaro2021quantum} regards as a test of the `quantum equivalence principle.' Now, we take it that  both clocks must be associated with registers if they are to be regarded as reference frames, and for the moment let us also take it that both clock systems are associated with \emph{macroscopic} registers $R_A, R_B$. We will also adopt the plausible assumption that the relation `is temporally localised relative to' is transitive and symmetric. Clearly it is also reflexive, and therefore it defines an equivalence relation which partitions events and objects into equivalence classes within which everything is temporally localised relative to everything else. 
Now suppose event $1$ is temporally localised relative to register $R_A$ and event $2$ is temporally localised relative to register $R_B$. Since registers $R_A$ and $R_B$ are macroscopic they will presumably be temporally localised relative to one another, as decoherence will very quickly destroy any coherent superpositions; it follows that event $1$ is temporally localised relative to event $2$, and vice versa. 
Thus while it certainly can be the case that some quantum event is not temporally localised relative to some reference frame, as we saw in section \ref{localisation}, it is less clear that one reference frame $R_1$ can be delocalised relative to a second reference frame $R_2$ - if we require that both reference frames are equipped with \emph{macroscopic} registers, then they will necessarily be localised with respect to one another, and therefore events which are localised relative to one reference frame will necessarily also be localised relative to the other. 

Of course this logic does not necessarily follow if we allow that the reference frames are associated  with \emph{quantum} registers, which may indeed be delocalised relative to one another. Arguably that is what is going on in ref \cite{cepollaro2021quantum}, where the internal degrees of freedom of the atomic clocks may in some sense be regarded as registers. But recall that the final-measurement interpretation requires that in order to be physically meaningful the information stored in the registers must  be stored in stable ongoing records, and certainly the easiest way to achieve that is to make the registers macroscopic. That said, it's possible that the registers could be quantum systems at the time of the generalized measurements and the information stored in them could subsequently be transferred into a macroscopic register, so the reference frames could be delocalised relative to one another at the time of the measurement and obtain definite states only later on. Because the relational variables describing events in frames which are not temporally localised relative to one another will typically not commute, it will not in general be possible to find a choice of basis for the final measurement such that over many repetitions of this setup both events are always recorded as occurring at a well-defined sharp time, but what we \emph{will} find is that if we repeat this setup many times, when we measure in a basis which reveals a sharp time for event $1$ then   the time for event $2$ will exhibit some spread, and when we measure in a basis which reveals a sharp time for event $2$ then the time for event $1$ will exhibit some spread.  Thus we conclude that within the final-measurement interpretation,  it is indeed possible to have two reference frames delocalised relative to one another as argued by refs \cite{2020qctloe,cepollaro2021quantum}, but in this case the reference frames in question must be set up very carefully such that their registers are quantum at the relevant time but the information about time localisation is subsequently transferred into macroscopic degrees of freedom.

This also suggests a way of expressing the quantum equivalence principle which allows it to have indirect operational significance without requiring measurements performed by actual observers to be in superposition relative to one another. That is to say, rather than expressing the  equivalence principle in terms of observations, observers or measurements, we should formulate it with respect to records in a register which are understood to be observed at a time later than they were recorded.  For example, rather than asserting that `\emph{The outcome of any local nongravitational experiment is independent of the position and velocity of the freely falling quantum reference frame in which it is performed}'\cite{cepollaro2021quantum} one could instead assert that the records expected to result from performing generalised measurements on local non-gravitational degrees of freedom within a IQRF are independent of the position and velocity of the freely falling IQRF.  This gives a clear (albeit indirect) operational meaning to the assertion of equivalence which may indeed be applied even in the case where the IQRFs in question are in superpositions relative to one another, meaning that under this construal the quantum equivalence principle can indeed be regarded as having  content that goes beyond the classical equivalence principle.

\section{Relational superposition  \label{rs} } 

We now move to an evaluation of one particularly intriguing  outcome of the IQRF research programme - the conclusion that   superposition is \emph{relational}. That is to say, a system may be in a superposition in some fixed basis relative to one reference frame but not another\cite{2019cov}. (Note that there are several other ways in which superposition might be said to be relative; we discuss how these types of relational superposition differ to the IQRF approach in appendix \ref{sup}, but here we will focus on the IQRF notion that superposition \emph{in a fixed basis} may be relative to the choice of internal quantum reference frame). Similarly, the IQRF approach also tells us that entanglement is relational - systems that are not entangled in one reference frame may appear to be entangled in another reference frame, or the amount of entanglement may change. The underlying explanation for this behaviour is the fact that tensor decompositions are not in general invariant under  quantum reference frame transformations\cite{delahamette2021perspectiveneutral}, so degrees of freedom may be shifted from one subsystem to another under such transformations: therefore each IQRF actually contains a different set of subsystems and thus inevitably these different subsystems also exhibit different amounts of superposition and entanglement.  
 
Now, it may be tempting to draw strong foundational conclusions from the fact that  superposition in a fixed basis is relative to an internal reference frame. For example, consider a laboratory containing a quantum particle which has been prepared in a superposition of position basis states. The IQRF framework tells us that systems don't have a quantum state relative to themselves, and hence by default the laboratory reference frame says that the laboratory has a well-defined position; but in the quantum reference frame of the \emph{particle} the laboratory is in a superposition of  position basis states. This would seem to have serious consequences for a range of interpretations of quantum mechanics, because many interpretational approaches are founded specifically on the conviction that macroscopic objects can't be in superpositions; for example, the de Broglie-Bohm interpretation\cite{holland1995quantum}, spontaneous collapse models\cite{tumulka2020relativistic} and gravitational collapse models\cite{Penrose1996OnGR} are all designed to avoid macroscopic superpositions. Of course, superposition is always relative to a basis, so we should be more specific:  what these models actually aim to do is avoid superpositions  in bases corresponding to macroscopically distinct measurement outcomes.  Thus these interpretations  can be jointly characterised by the mantra that `measurements have unique outcomes,' by contrast to Everett, QBist  and Copenhagen-style interpretations, which don't always insist on unique measurement outcomes. 
 
The notion that superpositions are relative to reference frames seems to pose a threat to this mantra, because it now appears that  there can be no fact of the matter about whether measurements have unique outcomes, since uniqueness of outcomes is defined only relative to a  reference frame. One might therefore begin to worry that  any interpretation of quantum mechanics which insists on unique measurement outcomes must break   some kind of covariance principle. On the plus side, one might begin to wonder if there is some novel interpretation of quantum mechanics lurking in this vicinity - perhaps the fact that no quantum system is ever in a superposition relative to itself is enough to explain why we experience measurements as having unique outcomes, even though from the perspective of other frames their outcomes may not be unique?  For example, Hardy seems to be suggesting something of this kind when he writes, `\emph{This offers some hope that macroscopic superpositions and entanglement based nonlocality are gauge artefacts and will not truly be present amongst the beables.}' \cite{hardy2019implementation}. However, an interpretation of quantum mechanics based on this observation would necessarily have some of the same problems as the Everett interpretation with regard to quantum probabilities. For suppose  we say that  an observer who measures a particle superposed relative to the lab reference frame ends up in a superposition relative to that reference frame, but she herself perceives herself as making a single definite observation, since she can't be in a superposition relative to her own reference frame. But then it seems hard   to say anything concrete about which measurement outcome the observer  sees, and in particular it's difficult to make sense of the probability distribution over her measurement outcomes, since whatever she does observe, it seems she must observe the same thing every time she performs this measurement.  Since we are not convinced that the Everettian probability problem can be solved, we also have doubts about whether these similar probability problems could be solved; so let us instead  try to determine how these assertions would fare   in the context of a single-world realist interpretation. 

To start with let us use the final-measurement approach as an example. Consider the case where a laboratory `sees' a particle as being in a superposition state relative to the position basis, whereas the particle `sees' the laboratory as being in a superposition state relative to the position basis. In the final-measurement approach, this means that there is no possible choice of measurement on the register at the end of time which will select a definite state both for the  laboratory in the  position basis and for the particle in the position basis during this period. (Another way to see this is to note that the relational observables associated with the position states for the particle and the laboratory are non-commuting). Thus far, the description looks entirely neutral between the particle and the laboratory, which might in some sense be considered to vindicate the claim that superposition is a relative matter, since it seems that either of the two descriptions is valid up until to the point far in the future where the final-measurement is actually performed and one of these options is selected.  

But in fact this is not correct, because we do not have freedom to select either a final measurement which resolves a definite state for the laboratory or a final measurement which resolves a definite state for the particle. For the final measurement can resolve a definite state for a system at a time only if there is a record of that state in the final state of the universe, and for quantum particles in what we typically refer to as `superpositions' there is no such record. Indeed, according to the usual way of using the term `superposition,' as soon as a record is made the system immediately ceases to be in a `superposition'  relative to the basis employed by the record - so for example, in the double-slit experiment, as soon as we perform a measurement determining which slit the particle passes through (which in the generalized measurement formalism would be modelled as an operation mapping the state of the particle in the `which-slit' basis to a register) that particle ceases to be in a superposition relative to the `which-slit' basis. Thus in the particle-laboratory case, there can't possibly be any choice of measurement on the register at the end of time which will select a definite state for the particle in the position basis, because by definition the position of the particle during this period is not written to any register which can be measured at the end of time.  

On the other hand, there are good reasons to suppose that a  measurement at the end of time will indeed single out a definite state for the laboratory in the position basis. This conclusion is   based not solely on the fact that in our experience laboratories are not usually found in superpositions of different positions, but also on what we know about the effect of decoherence\cite{sep-qm-decoherence} - the state of the laboratory at time $t$ will leave a large number of traces on its environment, which is to say we can always regard the state of the laboratory in the position basis as being recorded in some `register' which is measured at the end of time. Indeed, the state of the laboratory in the position basis will be recorded with a high degree of redundancy in the form of many different environmental traces, and therefore  it is more or less guaranteed that the final-measurement will resolve a definite position-basis state for the laboratory at time $t$.  So it is not really the case that the description where the laboratory is in a superposition relative to the  position basis is equally as valid as the description where the particle is in a superposition relative to the  position basis: it is true that both descriptions are mathematically possible ways of characterising the relation between these systems, but only the latter description properly captures the way in which the two systems are ultimately embedded in the ongoing records that ultimately determine the content of reality. 

Generalizing beyond the final-measurement approach, we emphasize that it is an interpretation-neutral fact that the behaviour of  macroscopic instruments in the laboratory   leaves a large number of traces on the environment which are carried forward into the future, whereas the behaviour of a superposed quantum particle leaves no traces,  and therefore the behaviour of the laboratory instruments has an ongoing impact on the history of the world whereas the behaviour of the particle is irrelevant to future history. Thus any single-world realist interpretation will most likely have to agree with our conclusion that the behaviour of the laboratory is determinate in a way that the behaviour of the quantum particle is not - e.g. because wavefunction collapses  ensure that the macroscopic objects in the laboratory are in definite states, or because macroscopic objects supervene on the de Broglie-Bohm variables which always have a definite state.   Indeed, even branching approaches such as the Everett interpretation will at least  acknowledge that the behaviour of the macroscopic laboratory instruments is necessarily a stable, well-established feature of each individual branch, whereas most branches will be largely unaffected by the behaviour of the particle. Therefore we contend that regardless of interpretation it is not correct to regard the superposed laboratory description and the superposed particle description as equally valid representations of reality: instantaneously there may be no mathematical difference between them, but the description where the laboratory has a definite state is much better fit when we consider the way in which this instantaneous description slots into a temporally extended account of the contents of reality. 

The same points apply to the relativity of entanglement and the relativity of tensor product decompositions. Different quantum reference frames will yield different decompositions into subsystems and thus different descriptions in terms of entanglement and superposition, but that does not mean that all of these decompositions are equally useful: some subsystem decompositions capture  more of the dynamically and operationally relevant features, and therefore even if we believe that no subsystem decomposition is explicitly favoured by the laws of nature, nonetheless certain decompositions will be `special' at least at an emergent level. Indeed, this is true even in classical physics. In the classical world we typically identify subsystems in terms of neighbouring particles which are interacting closely, so that at some level of approximation the subsystems can be treated as autonomous dynamical units under coarse-graining; mathematically speaking one could always choose to identity a `subsystem' made up of two non-interacting particles on opposite sides of the system, but this would lead to a complicated and unilluminating dynamics. Similarly in the quantum case, if our aim is to treat subsystems as autonomous dynamical units, it makes sense to use subsystem decompositions relativized to reference frames associated with macroscopic objects, because those are the decompositions which will likely be dynamically favoured.   So even if preferred subsystem decompositions  do not appear explicitly in the most fundamental laws of nature,  nonetheless we have good reasons to give special emphasis to subsystem decompositions relative to macroscopic reference frames.

These observations suggest a different way of thinking about the contribution that the IQRF programme can make to foundational problems. The IQRF programme started out by showing us how to do physics without assuming a predefined background frame, and thus it treats  all reference frames are formally equivalent, but  presumably it will ultimately go on to show that when dynamical considerations are taken into account certain   reference frames come to have special significance, such that our usual modes of classical description can be recovered. So rather than simply showing that superposition, entanglement and subsystem decompositions are irreducibly relational, the framework can be regarded as showing  that our standard `absolute' notions of superposition, entanglement and subsystem decomposition do not need to be predicated on an `absolute' background relative to which the superpositions, entanglement and subsystems are defined: the de facto `absolute' reference frame constituted by the macroscopic objects we see around us can emerge out of entirely relational facts. In this sense, rather than undermining the usual notion that we can think of laboratories as having definite states in a way that particles in superpositions do not, the IQRF programme actually puts that notion on more solid ground by demonstrating that we can make sense of such a notion even if we believe that all physical quantities should ultimately be relational.

\section{Conclusion}

We have argued that the final-measurement approach offers a particularly appealing interpretation of the PW and IQRF research programmes, as it preserves all their mathematical results and allows us to do justice to the claim that the PW formalism provides an operational account of the nature of time. Clearly any putative account of time must at least say something about the nature of temporal experience, and the final-measurement interpretation  shows that the PW formalism is able to do so by reducing temporal facts to correlations exhibited in the register readings selected by the final-measurement. Moreover,  it is reassuring to see that there appears to be a workable single-world interpretation of the IQRF and PW approaches. 

We then asked   whether the foundational conclusions that can be drawn from the PW and IQRF frameworks are sensitive to a choice of interpretation. We noted that the    `quantum reference frames' invoked in the IQRF framework are not automatically linked with the sorts of actual or possible operational procedures that typically define reference frames,  so it is necessary to adopt some  interpretation  before we can determine if they  constitute reference frames in the usual sense - for example, without a clear link to actual or possible operational procedures, such reference frames might not be suitable to define an `equivalence principle,' insofar as the purpose of that equivalence principle is to  secure the local accessibility of the spacetime metric in terms of possible measurement operations.  Similarly, we have offered an important caveat to the relativity of superposition within the IQRF framework:  it is true that instantaneously we can rewrite the description of a  particle in a superposition state relative to a laboratory in such a way as to regard the laboratory as being in a superposition state relative to the particle, but in many cases it is most perspicacious to understand superposition in a way that takes into account the role it plays in ongoing dynamical evolution and the formation of memories and records. In both cases, we argued that the  operational content of the IQRF framework appears only in the limit as the size of at least one reference system becomes large, and that therefore the IQRF framework has an important role to play in showing how our macroscopic reference frames can emerge out of wholly relational facts. 

Independently of the particular interpretational stance we have adopted here, we also want to make  a broader methodological point about the importance of interpretation to the project of foundational research.  Given the continuing debate over the correct interpretation of quantum mechanics, it may seem very tempting to proceed with studying the foundational consequences of the well-understood framework of unitary quantum mechanics without worrying about the measurement problem. But the important conceptual issues surrounding topics like the equivalence principle and superposition can't simply be reduced to technical questions within a mathematical framework:  both the equivalence principle and the notion of superposition have their applicability partly in virtue of the way in which they are tied into  the experiences of actual or possible observers, and therefore in order to say anything  about the consequences of the PW formalism and the IQRF framework for these sorts of topics, it is necessary to adopt some interpretation which allows us to understand the link between the mathematical formalism and actual or possible observations. And as we have seen in this article, the choice of interpretation does indeed have a substantial impact on the foundational conclusions that we can draw from the formalism. Indeed  this is likely to be the case for any attempt to `quantize' the notion of a reference frame: in both special and general relativity, reference frames are closely tied to observations and measurements, and thus it would be surprising if we could come up with a fully satisfactory quantum analogue of a `reference frame' without paying some attention to the thorny conceptual issues surrounding measurement in quantum mechanics. Of course this is not to say that we can't do these sorts of theoretical investigations without first solving the measurement problem, but we should keep in mind that the implications of those theoretical investigations will not usually be capable of being understood in a completely interpretation-neutral way. 

We would also like to emphasize that paying attention to questions of interpretation in the context of novel emerging research programmes has the potential to inject new life into what some physicists regard as a stale and futile debate over the interpretation of quantum mechanics. For example, we saw in this article that considering various proposals for the implementation of two-time measurements within the PW formalism seems to point us in the direction of the final-measurement interpretation originally advocated by Kent, and the observation that in this framework it makes sense to defer all measurements until `the end of time' may be regarded as undermining the Everettian claim that we are forced to accept multiple worlds if we want to maintain unitary quantum evolution. It is to be hoped that continued enquiries into the interpretation of other research programmes beyond standard textbook quantum mechanics may bring further clarity to these questions. 

We close by observing that the conclusions reached in this paper may also have interesting consequences for ongoing research in quantum gravity, since the Wheeler DeWitt equation plays an important role in quantum gravity and the PW formalism has sometimes been suggested as a solution to quantum gravity's famous `problem of time.' In particular, Loop Quantum Gravity was initially based on a partial solution to the WDW equation\cite{2015seqg}; the field has since developed other methods, such as spin foams, but conceptually we are still dealing with the problem of extracting information from a universal wavefunction as in the WDW equation. Similarly, the covariant path integral approach to quantum gravity also makes use of the WDW equation\cite{2015seqg}. Hence, although we have not addressed quantum gravity explicitly in this paper, some of our conclusions may translate to various quantum gravity approaches. In particular, geometrodynamics\cite{Anderson2004GeometrodynamicsSO}, loop quantum gravity\cite{Rovelli2008} and string theory\cite{Blau2009} are all noticeably lacking anything that looks like   a wavefunction collapse, and therefore one might be inclined to conclude that they will make sense only in the context of  the Everett interpretation, a Copenhagen or neo-Copenhagen interpretation, or some other such interpretation which does not accommodate unique measurement outcomes. However, the evident suitability of the final-measurement picture as an interpretative approach to the PW formalism may indicate a possible route to a single-world realist interpretation of LQC and other quantum gravity formalisms: we can treat the wavefunction unitarily from every possible reference frame, thus building up spacetime according to our preferred method, and then in the limit as all evolution becomes trivial we simply apply a `final measurement' to select one possible spacetime out of all the different possible spacetimes. Of course it remains to be seen if the details of this approach can be worked out coherently in the context of specific quantum gravity models, and this option must be compared to alternatives such as the Bohmian formulation of quantum gravity, but insofar as the final-measurement picture seems to emerge naturally from the existing formalism it seems a compelling option for those who would like to see a single-world realist take on quantum gravity. 

\appendix

\section{Other Relational Formalisms \label{others}} 
   
 In addition to the PW formalism, there are several other formal frameworks which enable us to obtain relative states from a global state in a similar manner. One approach uses relational Dirac observables\cite{Gambini:2008ke}, which encode how one observable evolves relative to another along the flow generated by the Hamiltonian constraint. Relational Dirac observables are often invoked in the context of classical Hamiltonian approaches, but we can obtain corresponding variables in the quantum case by first constructing variables in the kinematical Hilbert space which describe some system property conditional on some reference frame being in some orientation, and then mapping them to Dirac observables on the physical Hilbert space using a G-twirl\cite{delahamette2021perspectiveneutral}. We also have the relational Heisenberg picture\cite{2021trinity}, where we begin by trivializing the constraint so it acts only on the degrees of freedom of the reference frame (e.g. the clock) - that is, if T is the trivialization map and C the Hamiltonian constraint operator, we have $T H T^{-1} = (H_C - \epsilon) \otimes I_S$ for any real $\epsilon$. This transforms physical states into product states with a fixed redundant clock factor, and we then project onto the classical gauge-fixing condition to remove the redundant reference frame degrees of freedom. These various approaches differ in their technical details but all present a roughly similar picture in which  temporal notions are understood in terms of states of the universe relative to some physical reference frame.  Indeed ref \cite{2021trinity} demonstrates that the PW formalism, the relational Dirac observable formalism, and the relational Heisenberg formalism are all in fact equivalent. Ref \cite{2021trinity} also demonstrates that in all three cases, the procedure for changing temporal reference frames amounts to transforming from a state relative to some system $A$ to a global state which encodes all the possible relative states and then finally applying a reduction procedure to arrive at the state relative to some other system $B$. This demonstrates the vital importance of the global perspective-neutral picture over and above all the individual relative-state pictures, and indeed ref \cite{2019switch} argues that it offers a new way of interpreting the notion of the `quantum state of the universe' as an encoding of all these relative states. 

We should also distinguish between the IQRF approach and the related  reference frame research programme in quantum information\cite{RevModPhys.79.555,PhysRevA.99.052315,Gour_2008}.  In  the QI context, reference frames typically come into play in the  operational setting where we have a pair of observers who wish to exchange information but who do not share any common reference frame, which is to say we have no guarantee that they will perform measurements in the same basis. The solution is to encode information into relational degrees of freedom- for example, we could encode a bit into two qubits with $0$ associated with the state $\frac{1}{\sqrt{2}} (| 00 \rangle + |11 \rangle) $ and $1$ associated with the state $\frac{1}{\sqrt{2}} (| 01 \rangle + |10 \rangle) $, because the  answer to the question `are the states of the two qubits the same or different?' will be the same regardless of what basis we measure in. The key difference between these two approaches is that the QI formalism employs observers who are external to the quantum system in question, rather than seeking to define reference frames from within the system as the IQRF programme does: hence in the QI approach the set of states which are regarded as being legitimate in the absence of an external reference frames are the incoherently group-averaged states, rather than the coherently group-averaged states as in the IQRF approach,  reflecting the fact that the QI approach uses external observers who cannot themselves be entangled with parts of the system in question\cite{2021mmetc}. The QI approach therefore poses no particular interpretational challenges over and above the usual measurement problem,  as all interpretations of quantum mechanics will make the same predictions for the results of the measurements performed by these external observers. Whereas the correct way to model observers as internal features of a quantum system remains highly controversial\cite{2018qtcc,2020unqtcc},  and therefore the internal IQRF approach lies right at the meeting point of a number of thorny conceptual issues in quantum foundations. We therefore focus in this article on the internal reference frame programme rather than the related QI approach. 

\section{Final-measurement interpretation \label{fm}}

 The final-measurement interpretation allows us to  say more about the notion that time is just gauge. It is still true in this picture that specifying a state at a time or a single relative state is enough to determine the whole universal quantum state, so in that sense it remains true that time is a form of gauge, at least with respect to the universal quantum state. But in the final-measurement picture the relative states don't really refer to anything physically real, so the fact that they are all related by gauge transformations may seem less surprising: we simply have a choice of gauge in which to express the  probability rule which is used to determine the appropriate probability distribution for the final measurement. On the other hand, if we now look at the actual course of history selected by this measurement, we will see apparent indeterminism corresponding to the probabilistic results of quantum measurements, so it is \emph{not} true for this actual course of history that the state at one time is sufficient to determine the state at all other times. Therefore time is not merely gauge for the \emph{real} timeline as defined by the register readings - genuinely new information comes into being at different times along this timeline. This suggests that the problem of time in quantum gravity is to some extent an artefact of an approach which reifies the universal quantum state whilst sweeping the measurement problem under the rug: if one includes some actual measurement results the picture looks very different and there is less reason to be concerned about `timelessness.' Note that this consequence is not unique to the final-measurement interpretation: as observed by Callender and Weingard\cite{CWxx}, `\emph{contrary to the static situation described at the outset, a Bohmian approach to cosmology admits nontrivial evolution of the dynamical variables.}' So it seems that quite generally, as soon as we we add to the universal quantum state some variables describing what \emph{actually} occurs, we  no longer have the problem of timelessness. Indeed, from the point of view of approaches like the final-measurement interpretation and the Bohmian interpretation, the universal wavefunction is in a sense simply describing a set of counterfactual possibilities, and thus its timelessness is not at all surprising -  for presumably one would not expect a set of counterfactual possibilities to undergo change or evolution in any case.

Given that the final measurement on the register will necessarily take the state outside the physical Hilbert space, one might perhaps have concerns that this interpretation would break the symmetries which have been used to set up the physical Hilbert space.   And indeed, it is true that if we define a post-measurement state in the usual way this state will no longer belong to the physical Hilbert space; but by definition the final-measurement occurs after all meaningful dynamical evolution is complete, so we are not required to actually perform a projection into a post-measurement state since nothing happens after the final measurement.   The constraint quantisation ensures  that the probability distribution defined over possible register readings by the quantum state is invariant under diffeomorphisms and global time translations, and therefore the process of selecting one set of register readings according to that probability distribution is also necessarily invariant under diffeomorphisms and global time translations. Moreover, the constraints also ensure that the `course of history' thus selected is defined only up to diffeomorphisms and global time translations, which ensures that we are not `double counting'  - i.e. two courses of history related by a global time translation are not treated as distinct possibilities by the  probability distribution, they are assigned the some probability and   the  sum over histories required to normalise the probability distribution counts this probability only once. So the final-measurement is compatible with the  symmetries used to define the quantised theory, and therefore the final-measurement approach does not violate the founding idea of the IQRF approach that all physical quantities are ultimately relational: the course of history that we select in the final measurement is definite and observer-independent, but it does not presuppose an absolute background spacetime.

\section{Penrose \label{Penrose}}

Ref \cite{giacomini2021quantum} suggests an application for the quantum equivalence principle in the context of Penrose'se argument for gravitationally induced collapse\cite{Penrose1996OnGR}. That is, ref \cite{giacomini2021quantum} contends that Penrose's argument can be blocked by observing that he is using the wrong Equivalence Principle - he is working with the classical Equivalence Principle whereas he should in fact be using the Quantum Equivalence Principle. Now if this were correct, it would clearly be a counterexample to our worry that the quantum equivalence principle fails to have operationally meaningful content that goes beyond the classical SEP. But  let us  consider in more detail the way in which Penrose's argument uses the Equivalence principle. Penrose's fundamental concern is that in cases where we have a superposition of two different spacetimes there is no unique mapping from  the points of one spacetime to the points of another, since the principle of general covariance prevents us from assigning primitive identities to spacetime points which would identify them across branches of the wavefunction.  Penrose then worries that the global time-translation operator does not seem to be well-defined unless we can find a natural way of mapping the points of one superposed spacetime to points of the other, and therefore asks how we might  go about constructing such a map. This is where Penrose invokes the Principle of Equivalence - he notes that \emph{`In accordance with the principle of equivalence, it is the notion of free fall which is locally defined, so the most natural local identification between a local region of one space-time and a corresponding local region of the other would be that in which the free falls (i.e. spacetime geodesics) agree.'} 

Penrose does seem to be invoking a version of the SEP, but note that he does not appear to be insisting that the principle of equivalence will be violated if there does not exist a way of identifying the two spacetimes such that the geodesics agree. Rather, he is making use of the \emph{operational} construal of the principle of equivalence, which he understands as the statement that the metric can be accessed by means of local observables in the form of geodesic motions:  if this is accepted. presumably any sensible mapping  between the two spacetimes  should map the geodesic motions of one onto the geodesic motions of the other. Note that Penrose in fact already seems to be  employing `quantum reference frames' in some sense - that is, he seems to take it that the correct way to use the SEP in this situation is to apply it  \emph{separately within each branch of the wavefunction} to draw conclusions about our local access to the metric inside each branch. This application of the equivalence principle is entirely separate to his questions about whether there exists a way of mapping the locally accessible features of the metric of one onto the locally accessible features of the metric of the other. He does not say anywhere that the existence of such a mapping is required by the equivalence principle, or that he thinks there must be one unified Lorentzian frame of reference encapsulating both branches of the wavefunction - indeed, such a thing would be meaningless from the operational point of view where frames of reference are associated with actual or possible measurements, since measurements surveying two macroscopically distinct metrics must necessarily take place in one branch or another, so it would be nonsensical to postulate a frame of reference associated with measurements surveying both metrics at once. This indicates that Penrose's argument can't be blocked by simply altering the equivalence principle to apply separately within different branches of the wavefunction, because that is exactly how Penrose is already using it.

 Note that it is not our intention to claim that Penrose's argument is right. There are  other possible objections to it - for example, ref \cite{giacomini2021quantum}  also notes (correctly, in our view) that there is no obvious reason why it should be possible to define a universal time-translation operator which applies across all the different branches of a spacetime superposition, since we can simply allow that time flows at different rates within the different branches. But whether or not there exists such an operator is nothing to do with the principle of equivalence: the idea that the stability of a state depends on the existence of a well-defined time evolution operator is a further assumption of Penrose's argument which is a consequence of the standard quantum-mechanical definition of a stationary state, so it is actually an assumption that comes from quantum mechanics rather than general relativity, and therefore it is not related to  his use of the principle of equivalence to identify the local observables providing access to the metric.

\section{Relational superposition \label{sup}}

On top of the claims about relational superposition made in the IQRF research programme, there are several other ways in which superposition might be said to be relational. First,  of course superposition is relational in a trivial sense of being relative to a basis: for example, a qubit in the state $|0\rangle$ is not in a superposition relative to the computational basis $\{ | 0 \rangle , |1 \rangle \}$, but it is in a superposition relative to the Hadamard basis $\{ | + \rangle, | - \rangle \}$, since $|0 \rangle = \frac{1}{\sqrt{2}} ( | + \rangle + |- \rangle)$. In that sense, the assertion that a quantum system is `in a superposition' is meaningless without the specification of a basis. But the claims about the relativity of superposition in the IQRF research programme are stronger than that: for there we find that given a certain fixed basis, whether or not a system is in a superposition \emph{relative to that basis} is itself relative to a choice of reference frame. For example, in ref \cite{2019cov} we are given an example where relative to particle $C$, the particles $A$ and $B$ are in the entangled state $\int dx |x \rangle_A |x + X \rangle_B$ so neither of them has a sharp state in the position basis relative to $C$, but when we switch to $A$'s reference frame we find the state of $B$ is now sharp in the position basis. 

Second, relativistic effects can cause superpositions to look different from different reference frames in settings where frames are moving quickly relative to one another. For example, in ref \cite{pikovski2015time} it is noted that a Hamiltonian is defined with respect to a given slicing of space-time in equal-time surfaces, so observers in different references frames will assign different interaction Hamiltonians to a given experiment and as a result will come to different conclusions about the amount of superposition and/or entanglement present in a given physical situation. While these effects are very interesting and suggestive, they don't seem to require any major conceptual revolutions, as the `reference frames' in question are the ordinary kinds of reference frames commonly invoked within special relativity, and we are already very familiar with the idea that unusual frame-relative effects may occur when such frames are moving quickly  relative to one another. On the other hand, the relativity of superposition invoked by the IQRF research programme occurs even in the absence of relative motion, so it can't be understood in the same way as relativistic frame-dependence.

Third, we can also see frame-dependence of superposition and entanglement in the QI approach to reference frames based on the existence of an external reference frames; in this setting frame-dependence arises because the state of a system undergoes decoherence when we discard one reference frame during the switch to another reference frame\cite{RevModPhys.79.555,palmer2014changing}. `Reference frame' is here being understood in a similar way as in the IQRF reference frame, i.e. reference frames are associated with individual physical objects, and therefore much of what we are about to say regarding the relativity of superposition in the IQRF framework also apply here. However, it's important to note that the frame-dependent effects in the QI case can be understood as a consequence of a system (the initial reference frame) being discarded, whereas in the IQRF framework this explanation is not available since we are dealing with the whole universe at once and therefore nothing can be discarded.

 \section{Acknowledgements} 
 
 Thanks to Flaminia Giacomini and Philipp  Hoehn for their very helpful comments on a draft of this paper. 
 
 This publication was made possible through the support of the ID 61466 grant from the John Templeton Foundation, as part of the “The Quantum Information Structure of Spacetime (QISS)” Project (qiss.fr). The opinions expressed in this publication are those of the author  and do not necessarily reflect the views of the John Templeton Foundation.

 \bibliographystyle{unsrt}
 \bibliography{newlibrary12}{}

\begin{thebibliography}{10}

\bibitem{2020cqtd}
Alexander R.~H. Smith and Mehdi Ahmadi.
\newblock Quantum clocks observe classical and quantum time dilation.
\newblock {\em Nature Communications}, 11(1), Oct 2020.

\bibitem{2007mep}
Frank Hellmann, Mauricio Mondragon, Alejandro Perez, and Carlo Rovelli.
\newblock Multiple-event probability in general-relativistic quantum mechanics.
\newblock {\em Physical Review D}, 75(8), Apr 2007.

\bibitem{Kent}
A.~{Kent}.
\newblock {Solution to the Lorentzian quantum reality problem}.
\newblock {\em Phys Rev A}, 90(1):012107, July 2014.

\bibitem{2015KentL}
Adrian Kent.
\newblock Lorentzian quantum reality: postulates and toy models.
\newblock {\em Philosophical Transactions of the Royal Society A: Mathematical,
  Physical and Engineering Sciences}, 373(2047):20140241, Aug 2015.

\bibitem{2019cov}
Flaminia Giacomini, Esteban Castro-Ruiz, and Časlav Brukner.
\newblock Quantum mechanics and the covariance of physical laws in quantum
  reference frames.
\newblock {\em Nature Communications}, 10(1), Jan 2019.

\bibitem{2020qctloe}
Esteban Castro-Ruiz, Flaminia Giacomini, Alessio Belenchia, and Časlav
  Brukner.
\newblock Quantum clocks and the temporal localisability of events in the
  presence of gravitating quantum systems.
\newblock {\em Nature Communications}, 11(1), May 2020.

\bibitem{2020acop}
Augustin Vanrietvelde, Philipp~A. Hoehn, Flaminia Giacomini, and Esteban
  Castro-Ruiz.
\newblock A change of perspective: switching quantum reference frames via a
  perspective-neutral framework.
\newblock {\em Quantum}, 4:225, Jan 2020.

\bibitem{2020htsbrqc}
Philipp~A Höhn and Augustin Vanrietvelde.
\newblock How to switch between relational quantum clocks.
\newblock {\em New Journal of Physics}, 22(12):123048, Dec 2020.

\bibitem{giacomini2021quantum}
Flaminia Giacomini and Časlav Brukner.
\newblock Quantum superposition of spacetimes obeys einstein's equivalence
  principle, 2021.

\bibitem{dirac2019principles}
P.A.M. Dirac.
\newblock {\em The Principles of Quantum Mechanics}.
\newblock Lulu.com, 2019.

\bibitem{Kucha1993CanonicalQG}
Karel~V. Kuchař.
\newblock Canonical quantum gravity.
\newblock {\em arXiv: General Relativity and Quantum Cosmology}, 1993.

\bibitem{2015seqg}
Carlo Rovelli.
\newblock The strange equation of quantum gravity.
\newblock {\em Classical and Quantum Gravity}, 32(12):124005, Jun 2015.

\bibitem{hoehn2022quantum}
Philipp~A. Hoehn, Maximilian P.~E. Lock, Shadi~Ali Ahmad, Alexander R.~H.
  Smith, and Thomas~D. Galley.
\newblock Quantum relativity of subsystems, 2022.

\bibitem{2021mmetc}
Marius Krumm, Philipp~A. H\"{o}hn, and Markus~P. M\"{u}ller.
\newblock Quantum reference frame transformations as symmetries and the paradox
  of the third particle.
\newblock {\em Quantum}, 5:530, Aug 2021.

\bibitem{delahamette2021perspectiveneutral}
Anne-Catherine de~la Hamette, Thomas~D. Galley, Philipp~A. Hoehn, Leon
  Loveridge, and Markus~P. Mueller.
\newblock Perspective-neutral approach to quantum frame covariance for general
  symmetry groups, 2021.

\bibitem{2019switch}
Philipp Höhn.
\newblock Switching internal times and a new perspective on the “wave
  function of the universe”.
\newblock {\em Universe}, 5(5):116, May 2019.

\bibitem{2021gpr}
Veronika Baumann, Flavio~Del Santo, Alexander R.~H. Smith, Flaminia Giacomini,
  Esteban Castro-Ruiz, and Caslav Brukner.
\newblock Generalized probability rules from a timeless formulation of
  wigner’s friend scenarios.
\newblock {\em Quantum}, 5:524, Aug 2021.

\bibitem{PhysRevD.27.2885}
Don~N. Page and William~K. Wootters.
\newblock Evolution without evolution: Dynamics described by stationary
  observables.
\newblock {\em Phys. Rev. D}, 27:2885--2892, Jun 1983.

\bibitem{articleIsh}
Chris Isham.
\newblock Canonical quantum gravity and the problem of time.
\newblock {\em Integrable Systems, Quantum Groups, and Quantum Field Theories},
  11 1992.

\bibitem{doi:10.1142/S0218271811019347}
KAREL~V. KUCHAŘ.
\newblock Time and interpretations of quantum gravity.
\newblock {\em International Journal of Modern Physics D}, 20(supp01):3--86,
  2011.

\bibitem{2014fmqe}
Ekaterina Moreva, Giorgio Brida, Marco Gramegna, Vittorio Giovannetti, Lorenzo
  Maccone, and Marco Genovese.
\newblock Time from quantum entanglement: An experimental illustration.
\newblock {\em Physical Review A}, 89(5), May 2014.

\bibitem{PhysRevD.92.045033}
Vittorio Giovannetti, Seth Lloyd, and Lorenzo Maccone.
\newblock Quantum time.
\newblock {\em Phys. Rev. D}, 92:045033, Aug 2015.

\bibitem{PhysRevD.95.043510}
C.~Marletto and V.~Vedral.
\newblock Evolution without evolution and without ambiguities.
\newblock {\em Phys. Rev. D}, 95:043510, Feb 2017.

\bibitem{2021trinity}
Philipp H\"{o}hn, Alexander Smith, and Maximilian Lock.
\newblock Trinity of relational quantum dynamics.
\newblock {\em Physical Review D}, 104(6), Sep 2021.

\bibitem{2006poulin}
David Poulin.
\newblock Toy model for a relational formulation of quantum theory.
\newblock {\em International Journal of Theoretical Physics},
  45(7):1189–1215, Jun 2006.

\bibitem{vedralcewe}
V.~Vedral.
\newblock Classical evolution without evolution.
\newblock 2022.

\bibitem{2019quantizinginterac}
Alexander R.~H. Smith and Mehdi Ahmadi.
\newblock Quantizing time: Interacting clocks and systems.
\newblock {\em Quantum}, 3:160, Jul 2019.

\bibitem{Wallace}
David Wallace.
\newblock Everett and structure.
\newblock {\em Studies in History and Philosophy of Science Part B: Studies in
  History and Philosophy of Modern Physics}, 34(1):87--105, 2003.

\bibitem{AdlamEverett}
Emily Adlam.
\newblock {The Problem of Confirmation in the Everett Interpretation}.
\newblock {\em Studies in History and Philosophy of Science Part B: Studies in
  History and Philosophy of Modern Physics}, 47:21 -- 32, 2014.

\bibitem{Albert2010-ALBPIT}
David Albert.
\newblock Probability in the everett picture.
\newblock In Simon Saunders, Jonathan Barrett, Adrian Kent, and David Wallace,
  editors, {\em Many Worlds?: Everett, Quantum Theory \& Reality}. Oxford
  University Press, 2010.

\bibitem{kent2009world}
Adrian Kent.
\newblock One world versus many: the inadequacy of everettian accounts of
  evolution, probability, and scientific confirmation, 2009.

\bibitem{Dolby2004TheCP}
Carl~E. Dolby.
\newblock The conditional probability interpretation of the hamiltonian
  constraint.
\newblock {\em arXiv: General Relativity and Quantum Cosmology}, 2004.

\bibitem{craig2011consistent}
David Craig and Parampreet Singh.
\newblock Consistent histories in quantum cosmology.
\newblock {\em Foundations of Physics}, 41(3):371--379, 2011.

\bibitem{CWxx}
Craig Callender and Robert Weingard.
\newblock The bohmian model of quantum cosmology.
\newblock {\em PSA: Proceedings of the Biennial Meeting of the Philosophy of
  Science Association}, 1994:218--227, 1994.

\bibitem{1996consihit}
Fay Dowker and Adrian Kent.
\newblock On the consistent histories approach to quantum mechanics.
\newblock {\em Journal of Statistical Physics}, 82(5-6):1575–1646, Mar 1996.

\bibitem{2017}
Adrian Kent.
\newblock Quantum reality via late-time photodetection.
\newblock {\em Physical Review A}, 96(6), Dec 2017.

\bibitem{Durr2013}
Detlef D{\"u}rr, Sheldon Goldstein, and Nino Zangh{\`i}.
\newblock {\em Quantum Spacetime without Observers: Ontological Clarity and the
  Conceptual Foundations of Quantum Gravity}, pages 247--261.
\newblock Springer Berlin Heidelberg, Berlin, Heidelberg, 2013.

\bibitem{Einstein1905}
A.~{ Einstein}.
\newblock {On the electrodynamics of moving bodies}.
\newblock {\em {Annalen der Physik}}, 17:891 -- 921, 1905.

\bibitem{Scaranicloning}
V.~{Scarani}, S.~{Iblisdir}, N.~{Gisin}, and A.~{Ac{\'{\i}}n}.
\newblock {Quantum cloning}.
\newblock {\em Reviews of Modern Physics}, 77:1225--1256, October 2005.

\bibitem{sep-operationalism}
Hasok Chang.
\newblock {Operationalism}.
\newblock In Edward~N. Zalta, editor, {\em The {Stanford} Encyclopedia of
  Philosophy}. Metaphysics Research Lab, Stanford University, {F}all 2021
  edition, 2021.

\bibitem{2021qrft}
Marius Krumm, Philipp~A. H\"{o}hn, and Markus~P. M\"{u}ller.
\newblock Quantum reference frame transformations as symmetries and the paradox
  of the third particle.
\newblock {\em Quantum}, 5:530, Aug 2021.

\bibitem{hoehn2021internal}
Philipp~A. Hoehn, Marius Krumm, and Markus~P. Mueller.
\newblock Internal quantum reference frames for finite abelian groups, 2021.

\bibitem{giacomini2021einsteins}
Flaminia Giacomini and Časlav Brukner.
\newblock Einstein's equivalence principle for superpositions of gravitational
  fields, 2021.

\bibitem{2010dqmcw}
Elias Okon and Craig Callender.
\newblock Does quantum mechanics clash with the equivalence principle—and
  does it matter?
\newblock {\em European Journal for Philosophy of Science}, 1(1):133–145, Nov
  2010.

\bibitem{pittphilsci17709}
Dennis Lehmkuhl.
\newblock The equivalence principle(s), 2019.

\bibitem{synge1960relativity}
J.L. Synge.
\newblock {\em Relativity: The General Theory}.
\newblock Number v. 1 in North-Holland series in physics. North-Holland
  Publishing Company, 1960.

\bibitem{pittphilsci14589}
James Read, Harvey~R. Brown, and Dennis Lehmkuhl.
\newblock Two miracles of general relativity, March 2018.
\newblock Forthcoming in Studies in History and Philosophy of Modern Physics.

\bibitem{Einsteinmethods}
{Einstein, A.}
\newblock {Physical relativity space-time structure from a dynamical
  perspective}, 1920.

\bibitem{Einsteinletter}
A.~Einstein.
\newblock {Letter to Painlev\'{e}}, 1921.

\bibitem{Brownrelativity}
{Brown Harvey R.}
\newblock {\em {Fundamental Ideas and Methods of the Theory of Relativity,
  Presented in Their Development}}.
\newblock Clarendon Press ; Oxford University Press, 2005.

\bibitem{Knox2013-KNOESG}
Eleanor Knox.
\newblock Effective spacetime geometry.
\newblock {\em Studies in History and Philosophy of Science Part B: Studies in
  History and Philosophy of Modern Physics}, 44(3):346--356, 2013.

\bibitem{maudlin2012philosophy}
T.~Maudlin.
\newblock {\em Philosophy of Physics: Space and Time}.
\newblock Princeton Foundations of Contemporary Philosophy. 2012.

\bibitem{cepollaro2021quantum}
Carlo Cepollaro and Flaminia Giacomini.
\newblock Quantum generalisation of einstein's equivalence principle can be
  verified with entangled clocks as quantum reference frames, 2021.

\bibitem{albert1994quantum}
D.Z. Albert.
\newblock {\em Quantum Mechanics and Experience}.
\newblock Harvard University Press, 1994.

\bibitem{KNOX2013346}
Eleanor Knox.
\newblock Effective spacetime geometry.
\newblock {\em Studies in History and Philosophy of Science Part B: Studies in
  History and Philosophy of Modern Physics}, 44(3):346--356, 2013.

\bibitem{hardy2019implementation}
Lucien Hardy.
\newblock Implementation of the quantum equivalence principle, 2019.

\bibitem{Krestschmanngc}
E.~{Kretschmann}.
\newblock \"{U}ber den physikalischen sinn der relativit\"{a}tspostulate. a.
  einsteins neue und seine ursprüngliche relativit\"{a}tstheorie.
\newblock {\em Annalen der Physik}, page 575–614, 1917.

\bibitem{KochenSpecker}
Simon Kochen and E.P. Specker.
\newblock The problem of hidden variables in quantum mechanics.
\newblock In C.A. Hooker, editor, {\em The Logico-Algebraic Approach to Quantum
  Mechanics}, The University of Western Ontario Series in Philosophy of
  Science, pages 293--328. Springer Netherlands, 1975.

\bibitem{Spekkens}
R.~W. {Spekkens}.
\newblock {Contextuality for preparations, transformations, and unsharp
  measurements}.
\newblock {\em Phys Rev A}, 71(5):052108, May 2005.

\bibitem{holland1995quantum}
P.R. Holland.
\newblock {\em The Quantum Theory of Motion: An Account of the de Broglie-Bohm
  Causal Interpretation of Quantum Mechanics}.
\newblock Cambridge University Press, 1995.

\bibitem{tumulka2020relativistic}
Roderich Tumulka.
\newblock A relativistic grw flash process with interaction, 2020.

\bibitem{Penrose1996OnGR}
Roger Penrose.
\newblock On gravity's role in quantum state reduction.
\newblock {\em General Relativity and Gravitation}, 28:581--600, 1996.

\bibitem{sep-qm-decoherence}
Guido Bacciagaluppi.
\newblock {The Role of Decoherence in Quantum Mechanics}.
\newblock In Edward~N. Zalta, editor, {\em The {Stanford} Encyclopedia of
  Philosophy}. Metaphysics Research Lab, Stanford University, {F}all 2020
  edition, 2020.

\bibitem{Anderson2004GeometrodynamicsSO}
Edward Anderson.
\newblock Geometrodynamics: Spacetime or space ?
\newblock {\em arXiv: General Relativity and Quantum Cosmology}, 2004.

\bibitem{Rovelli2008}
Carlo Rovelli.
\newblock Loop quantum gravity.
\newblock {\em Living Reviews in Relativity}, 11(1):5, 2008.

\bibitem{Blau2009}
Matthias Blau and Stefan Theisen.
\newblock String theory as a theory of quantum gravity: a status report.
\newblock {\em General Relativity and Gravitation}, 41(4):743--755, 2009.

\bibitem{Gambini:2008ke}
Rodolfo Gambini, Rafael~A. Porto, Jorge Pullin, and Sebastian Torterolo.
\newblock {Conditional probabilities with Dirac observables and the problem of
  time in quantum gravity}.
\newblock {\em Phys. Rev. D}, 79:041501, 2009.

\bibitem{RevModPhys.79.555}
Stephen~D. Bartlett, Terry Rudolph, and Robert~W. Spekkens.
\newblock Reference frames, superselection rules, and quantum information.
\newblock {\em Rev. Mod. Phys.}, 79:555--609, Apr 2007.

\bibitem{PhysRevA.99.052315}
Alexander R.~H. Smith.
\newblock Communicating without shared reference frames.
\newblock {\em Phys. Rev. A}, 99:052315, May 2019.

\bibitem{Gour_2008}
Gilad Gour and Robert~W Spekkens.
\newblock The resource theory of quantum reference frames: manipulations and
  monotones.
\newblock {\em New Journal of Physics}, 10(3):033023, mar 2008.

\bibitem{2018qtcc}
Daniela Frauchiger and Renato Renner.
\newblock Quantum theory cannot consistently describe the use of itself.
\newblock {\em Nature Communications}, 9(1), Sep 2018.

\bibitem{2020unqtcc}
R.~E. Kastner.
\newblock Unitary-only quantum theory cannot consistently describe the use of
  itself: On the frauchiger–renner paradox.
\newblock {\em Foundations of Physics}, 50(5):441–456, Mar 2020.

\bibitem{pikovski2015time}
Igor Pikovski, Magdalena Zych, Fabio Costa, and Caslav Brukner.
\newblock Time dilation in quantum systems and decoherence: Questions and
  answers, 2015.

\bibitem{palmer2014changing}
Matthew~C Palmer, Florian Girelli, and Stephen~D Bartlett.
\newblock Changing quantum reference frames.
\newblock {\em Physical Review A}, 89(5):052121, 2014.

\end{thebibliography}

\end{document}